\documentclass[useAMS, twocolumn, usenatbib]{mn2e}
\usepackage{graphicx}
\usepackage{amsmath, amssymb}
\usepackage{natbib}

\begin{document}

\title{Evaporation and Accretion of Extrasolar Comets Following White Dwarf Kicks}
\author[Stone, Metzger, \& Loeb]{Nicholas Stone$^{1}$\thanks{E-mail: nstone@phys.columbia.edu}, Brian D.~Metzger$^{1}$, and Abraham Loeb$^{2}$\\
$^{1}$Columbia Astrophysics Laboratory, Columbia University, New York, NY, 10027\\
$^{2}$Astronomy Department, Harvard University, 60 Garden Street, Cambridge, MA 02138, USA}

\maketitle

\begin{abstract}
Several lines of observational evidence suggest that white dwarfs receive small birth kicks due to anisotropic mass loss.  If other stars possess extrasolar analogues to the Solar Oort cloud, the orbits of comets in such clouds will be scrambled by white dwarf natal kicks.  Although most comets will be unbound, some will be placed on low angular momentum orbits vulnerable to sublimation or tidal disruption.  The dusty debris from these comets will manifest itself as an IR excess temporarily visible around newborn white dwarfs; examples of such disks may already have been seen in the Helix Nebula, and around several other young white dwarfs.  Future observations with the {\it James Webb Space Telescope} may distinguish this hypothesis from alternatives such as a dynamically excited Kuiper Belt analogue.  Although competing hypotheses exist, the observation that $\gtrsim 15\%$ of young white dwarfs possess such disks, if interpreted as indeed being cometary in origin, provides indirect evidence that low mass gas giants (thought necessary to produce an Oort cloud) are common in the outer regions of extrasolar planetary systems.  Hydrogen abundances in the atmospheres of older white dwarfs can, if sufficiently low, also be used to place constraints on the joint parameter space of natal kicks and exo-Oort cloud models.  

\end{abstract}


\section{Introduction}
Metal absorption lines have been observed in white dwarf (WD) atmospheric spectra for many decades, but only recently has their importance for the study of exoplanetary systems become apparent.  Observations of metals in WD atmospheres were initially puzzling: WDs with either hydrogen (DA) or helium (DB) dominated atmospheres possess relatively short sedimentation timescales for heavier elements.  Given that early efforts to explain these observations through convective instability and dredge-up have been demonstrated unlikely, the atmospheric metal content of WDs must come from an external source.  Direct accretion from the interstellar medium fails to reproduce observed abundances by several orders of magnitude, and furthermore would predict metal pollution to be correlated with the galactocentric orbits of WDs in ways that are not observed \citep{Farihi+10}.  The general consensus at present is that these metals are actively accreted from the tidal disruption of small rocky bodies, likely asteroids that survived their host star's post-main sequence (MS) evolution and then were excited to high eccentricity by dynamical encounters with surviving planets \citep{DebSig02, Jura03}.  This assessment is strengthened by the detection of dusty debris disks around a significant fraction of WDs; these disks are only seen in association with metal-polluted WDs and are estimated to occur around $\sim 1 -3 \%$ of all single WDs \citep{Farihi+09}.

Metal pollution in white dwarfs has already been used to study the chemical composition and planetary architecture of rocky bodies in post-MS systems \citep{Jura+07}.  Less has been said on the volatile-rich bodies which likely form in most stellar systems beyond the ice line.  This is largely because measured abundances \citep{Klein+11, Gansic+12} have disfavored tidal disruption of hydrogen-rich objects as the primary source of metal pollution \citep{JurXu12}, although \citet{Farihi+13} provides a notable exception.  However, it is still plausible that tidal disruption of icy planets or planetesimals may contribute in a subdominant way to the total metal influx.  This class of sources could be subdominant at all times, or perhaps only in a time-averaged sense.  Past investigations of this possibility have focused on perturbations to surviving Oort cloud analogues from external sources such as passing stars \citep{DebSig02}, or scattering from planets that have survived post-main sequence evolution \citep{BonWya12}.  

We further explore the disruption and accretion of volatile-rich bodies in this paper, focusing on the orbital evolution of comets in exoplanetary Oort cloud analogues (hereafter OCAs).  In particular, anisotropic mass loss during the birth of a WD has the potential to radically alter OCA orbits, unbinding most \citep{ParAlc98} but putting a small fraction on nearly radial orbits.  Our paper is motivated by two puzzling observations concerning accretion on WDs.  
\begin{itemize}
\item Since the pioneering discovery of an IR excess around the central WD of the Helix nebula \citep{Su+07}, subsequent observations have established that $\sim 15\%$ of very young WDs possess IR excesses that can be explained by large-scale debris disks \citep{Chu+11}.  The young WDs referred to here generally have effective temperatures $\gtrsim 10^5~{\rm K}$ and ages $\lesssim 1~{\rm Myr}$, and often still have associated planetary nebulae.  Their dust disks have a large inferred radial extent, $\sim 10-100~{\rm AU}$, in stark contrast with the $<0.01~{\rm AU}$ disks that are observed around older DAZ or DBZ WDs.  
\item Many WDs are extremely hydrogen-deficient, with less H in their atmospheres ($M_{\rm H} < 10^{-15}M_{\odot}$) than would be delivered by a single, modestly sized comet \citep{Berger+11}.  This fact has been used in the past to constrain the density of interstellar comets \citep{Jura11}, but has not so far been applied to post-main sequence dynamics of OCAs.
\end{itemize}
The evaporation of a significant fraction of an OCA's mass following a natal WD kick would have clear implications for the first of these observations, and would be strongly constrained by the second.  However, evaporation of comets is not the only explanation for IR excesses seen around young WDs.  Debris disks around hot young WDs have been explained by enhanced collisions \citep{Su+07} among analogues to Kuiper Belt Objects (KBOs), driven by orbital perturbations from surviving gas giant exoplanets \citep{Dong+10}.  However, such a mechanism both requires survival of an appropriate exoplanet and of a sizeable population of KBO analogues.  The latter of these assumptions may be questionable.  An alternative explanation for these excesses is the formation of a dust-rich disk due to binary interactions with the AGB wind \citep{Biliko+12}, in analogy with the hot, dusty disks seen around post-AGB stars.  This possibility has gained credibility with the recent discovery of binary companions around a majority of young WDs with IR excesses \citep{Clayto+14}; we discuss both of these alternate hypotheses in \S 5.

We emphasize that the well-studied IR excesses seen around older, metal-polluted WDs are {\it not} the focus of this paper.  We are instead concerned with the poorly understood IR excesses seen around newborn WDs.  The dynamical model we introduce for the response of an OCA to natal WD kicks is meant to produce large amounts of dust on $\sim 10~{\rm AU}$ scales shortly after the birth of a WD, and has no connection to the $\lesssim 0.01~{\rm AU}$ size dusty disks seen in a subpopulation of DAZ/DBZ stars.

The following sections explore the orbital dynamics, destruction, debris evolution, and observability of this deeply plunging subpopulation of comets.  Specifically, in \S 2 we produce Monte Carlo samples of OCA objects (\S 2.1), evolve their orbital elements in response to both gradual and impulsive mass loss (\S 2.2), and verify the validity of the impulsive approximation (\S 2.3).  We next consider the fate of deeply plunging comets (\S 3), which can be destroyed either through tidal disruption (\S 3.1), or, more likely, through sublimation (\S 3.2).  The properties of the cometary debris are considered in \S 3.3.  In \S 4 we model the evolution and observable properties of the gaseous and solid-state debris disks formed from the leftovers of comet sublimation.  \S 5 provides a general discussion of the observational implications for OCAs around newborn WDs.  These include observable (and perhaps already observed) debris disks (\S 5.1), implications for WD natal kicks (\S 5.2), hydrogen abundances in tension with the most hydrogen-depleted WDs observed (\S 5.3), and surviving exoplanetary systems (\S 5.4).  We summarize our conclusions in \S 6.

We pay special attention to the two sets of observations mentioned above, focusing on both the observability of the solid state debris, and the total H accretion onto the WD.  As we will show, cometary debris disks around young WDs are promising sources for the {\it James Webb Space Telescope} (JWST).  The properties of such disks provide a probe of the properties of extrasolar Oort clouds and hence, indirectly, of the architectures of extrasolar planetary systems necessary to produce such a reservoir of icy bodies.  

\section{Response of Cometary Orbits to Mass Loss}
\label{sec:response}

As an AGB star sheds its envelope, the orbits of planets and planetesimals evolve in response.  Purely axisymmetric mass loss (that is also reflection-symmetric about the equatorial plane) increases the energy of surrounding orbits, while leaving all other orbital elements unchanged.  However, deviations from axisymmetry (or reflection symmetry) enable orbital eccentricities to increase.  Likewise, the rate of mass loss defines three qualitatively different regimes: orbits close to the star evolve adiabatically, orbits far from the star see the mass loss as impulsive, and in between lies a more complicated transition regime.

Non-axisymmetric mass loss may occur during this stage, for example due to MHD instabilities in AGB winds arising from magnetic cool spots on the stellar surface \citep{ThiHey10}.  Dust-driven AGB superwinds account for the plurality of post-main sequence mass loss for WDs with zero-age main sequence masses $M_{\rm ZAMS} \gtrsim 1.5M_{\odot}$, and contribute significantly to total mass loss for $M_{\rm ZAMS} \gtrsim 1.2 M_{\odot}$ \citep{Wachte+02}.   The extremely short duration of the superwind phase ($t_{\rm SW} \sim 3\times 10^4~{\rm yr}$) renders it the only plausible epoch capable of accounting for the (sometimes disputed) dynamical evidence for WD kicks observed in globular and open clusters (see \S 5.2). 

Equating $t_{\rm SW}$ to an orbital timescale gives a characteristic semimajor axis
\begin{equation}
a_{\rm imp}= 814~{\rm AU} \left(\frac{t_{\rm SW}}{3\times 10^4~{\rm yr}} \right)^{2/3} \left( \frac{M_{\rm WD}}{0.6M_{\odot}} \right)^{1/3}
\end{equation}
where the transition from adiabatic to impulsive mass loss occurs.  The adiabatic regime of mass loss is clearly the relevant one for planets that survive the AGB stage of their parent stars.  Although the Solar Kuiper Belt is located at a distance $a \sim 50~{\rm AU} \ll a_{\rm imp}$, the eccentricity of objects in extrasolar Kuiper Belt analogues will still evolve adiabatically \citep{Veras+13}, possibly to large values.  However, as we discuss in \S\ref{sec:dustydisk}, the ability of ice-dominated Kuiper Belt analogues to survive post-main sequence stellar evolution is highly questionable \citep{Stern+90, Melnic+01}.

In contrast to planets and the Kuiper Belt, the majority of the Solar Oort cloud is concentrated at radii $\gtrsim 1000$ AU $\gtrsim a_{\rm imp}$, placing the response of OCAs to AGB mass loss into the impulsive regime to first order.  As a result, OCA objects whose velocity vectors align with $\vec{v}_{\rm k}$, the kick velocity vector of the WD, can find their post-kick angular momentum decreased to very low values.  In this section, we use Monte Carlo simulations to explore the parameter space of plausible extrasolar Oort clouds and WD kicks, paying special attention to the subpopulation of comets placed onto deeply plunging orbits.

\subsection{Extrasolar Oort Clouds}

Although observations of debris disks around young stars provide some evidence for extrasolar comets \citep{Ferlet+87, Beust+90, Melnic+01, Rodiga+14}, it is unclear whether the progenitors of these disks originate in a coplanar, KBO-like configuration, or in a more spherical OCA.  Due to our limited information about the planetary dynamics of these systems, little can be said about the masses or radial distribution of their extrasolar comet reservoirs.  For these reasons our models of OCAs are based on our (limited) knowledge of the Oort cloud surrounding our own Solar System.

The Solar Oort cloud is a loosely bound, collisionless collection of icy bodies; based on observations of long-period comets, it is estimated to contain $\sim 10^{12}$ objects over a kilometer in size \citep{Weissman96}.  The Oort cloud is roughly spherically symmetric, with an inner boundary of $R_{\rm in} \approx 10^3~{\rm AU}$ set during its formation and modified by the Sun's orbit through the galaxy \citep{Kai+11}.  Often an inner Oort cloud beginning at this radius is distinguished from an outer Oort cloud starting at $\sim 10^4~{\rm AU}$ \citep{Fernan97}.  Mass estimates for the inner Oort cloud, the region most relevant for this paper, are highly uncertain because of the limited influence of the Galactic tide in this region \citep{Dones+04} and its consequent underproduction of observed long-period comets.  Theoretically, the inner regions of the Oort cloud are less thermalized in both angles and eccentricity than the outer regions, with a bias toward radial orbits somewhat aligned with the plane of the ecliptic.  

The total estimated mass of the entire cloud ranges between $2M_{\oplus}$ and $60 M_{\oplus}$ \citep{Weissman96, KaiQui08, Kai+11}, although the most recent dynamical work on this subject has favored lower estimates for the total Oort cloud mass, $\sim 7 M_{\oplus}$ \citep{KaiQui09}.  The Oort cloud is truncated by both the Galactic tide and encounters with passing stars, setting an outer edge $R_{\rm out} \approx 5-20\times 10^4 ~{\rm AU}$ \citep{Duncan+87}.  Although the detailed formation of the Oort cloud is not fully understood, a schematic picture is sketched in Appendix A.  In this simplified analytic model \citep{Tremai93}, a coplanar belt of planetesimals whose orbits cross those of large planets are perturbed to more weakly bound energies, and are eventually isotropized by the galactic tide.

More detailed properties of the Solar Oort cloud are predicted by simulations of planetesimal scattering off gas giants during the early history of the Solar System.  These simulations typically predict that OCA objects possess a power-law distribution of semimajor axes $n(a) \propto a^{-\gamma}$ with $\gamma \sim 2-4$, with numerical formation simulations generally finding $\gamma \approx 3.5$ \citep{Duncan+87, WieTre99, Brasse+06, KaiQui08}.

Our fiducial model assumes that cometary orbits possess an $a^{-3.5}$ semimajor axis distribution and a spherically symmetric distribution of inclinations, nodal angles, and lines of pericenter.  We take $R_{\rm in}=1000~{\rm AU}$ and $R_{\rm out}=5\times10^4~{\rm AU}$.  For each value of $a$, we sample from a modified thermal distribution of eccentricities.  The standard thermal distribution, $P(e)=2e$, is truncated at high $e$ so that no pre-kick pericenters below $500~{\rm AU}$ are sampled.  This is the rough pericenter value for which our approximation of impulsivity qualitatively breaks down (\S \ref{sec:nonImpulsive}), and is only a factor $\approx 2$ greater than the ice line during the star's AGB phase (\S \ref{sec:evap}).  We take a final WD mass of $M_{\rm WD}=0.6M_\odot$, but assume that it has lost half of its mass prior to the birth kick (we alter the $a$ and $e$ distributions to be consistent with this adiabatic and symmetric mass loss).  We also consider non-fiducial models that vary the assumption of spherical symmetry, $\gamma$, and $R_{\rm in}$ (our results are not sensitive to $R_{\rm out}$).  

\subsection{The Impulsive Limit}
\label{sec:impulse}

Having generated a large (N = 10$^{6}$) Monte Carlo sample of pre-kick comet orbits using the above prescriptions, these orbits are evolved in response to stellar mass loss off the main sequence.  This section considers the analytically tractable regime of impulsive mass loss, leaving the more general case to the next section. 

Orbital elements generated through Monte Carlo sampling are converted into six-dimensional Cartesian coordinates in position and velocity space.  These coordinates are then translated evenly in velocity space to account for the WD kick, before being converted back to standard orbital elements.  The transformations used by this procedure can be found in most celestial mechanics textbooks and are reviewed in Appendix B.  This procedure is used to explore the effect of different combinations of kick velocity $v_{\rm k}$ and Oort cloud parameters $\{ R_{\rm in}, R_{\rm out}, \gamma\}$ on the distribution of post-kick cometary properties.    

\begin{figure}
\includegraphics[width=85mm]{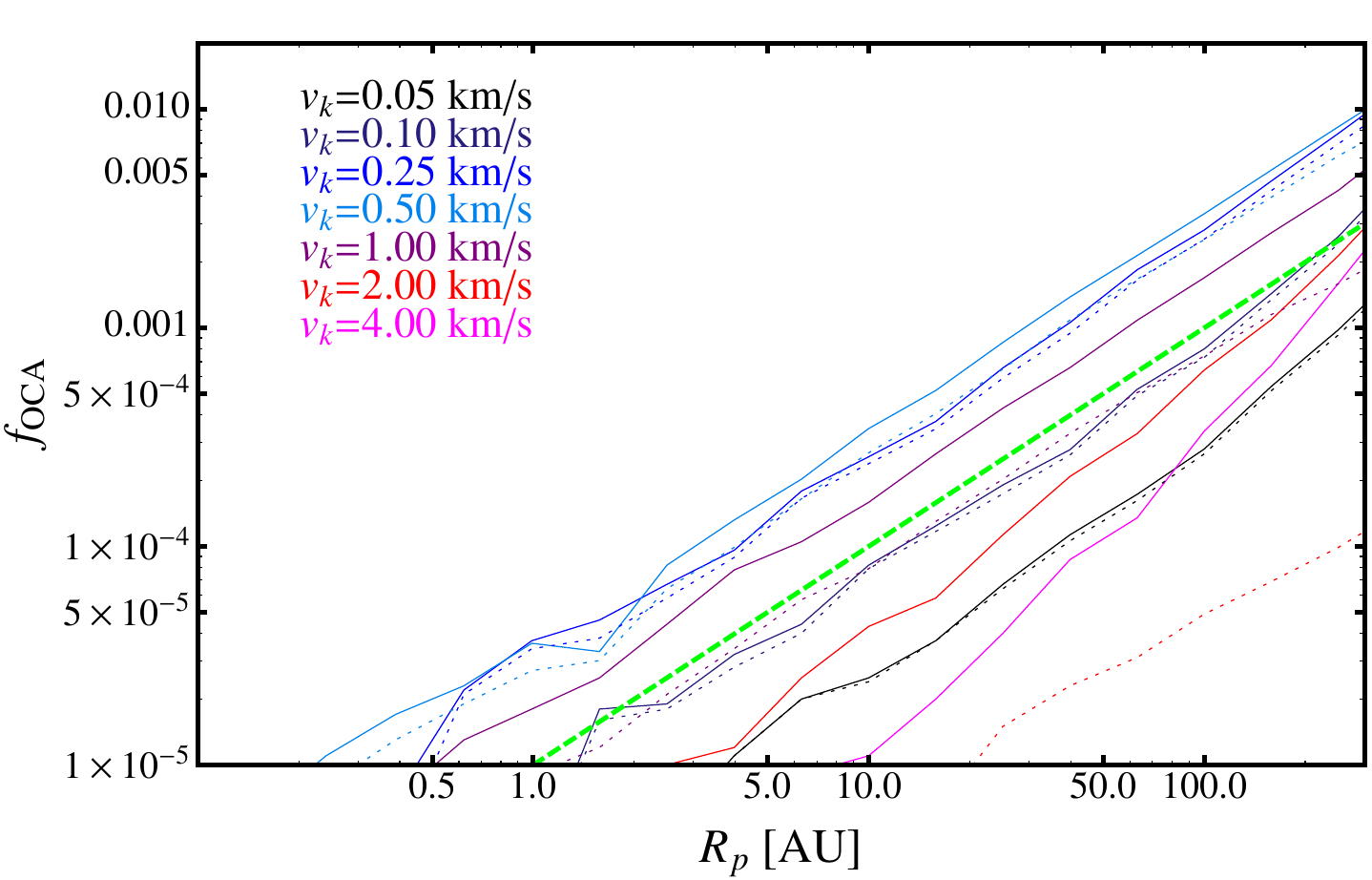}
\caption{Fraction of the mass of the OCA, $f_{\rm OCA}={\rm d}N/{\rm dln}R_{\rm p}$, deposited on orbits with pericenter radius $R_{\rm p}$ following the natal white dwarf kick.  Distributions shown are calculated for kick velocities $v_{\rm k} =$ 0.05 ({\it black}), 0.1 ({\it dark blue}), 0.25 ({\it blue}), 0.5 ({\it cyan}), 1 ({\it purple}), 2 ({\it red}), and 4 ({\it magnenta}) ${\rm km~s}^{-1}$, respectively.  The initial properties of the OCA adopted are motivated by models of the Solar Oort cloud: inner radius $R_{\rm in}=1000 ~{\rm AU}$, outer radius $R_{\rm out}=5\times 10^4~{\rm AU}$, and density profile $n(a)\propto a^{-\gamma}$ for $\gamma = 3.5$.  Solid lines represent all comets, while dotted lines only include the fraction of comets that remain bound to the WD after the kick.  The dashed green line, shown for reference, corresponds to an equal distribution of comets per linear pericenter distance.}
\label{periBoth}
\end{figure}

\begin{figure}
\includegraphics[width=85mm]{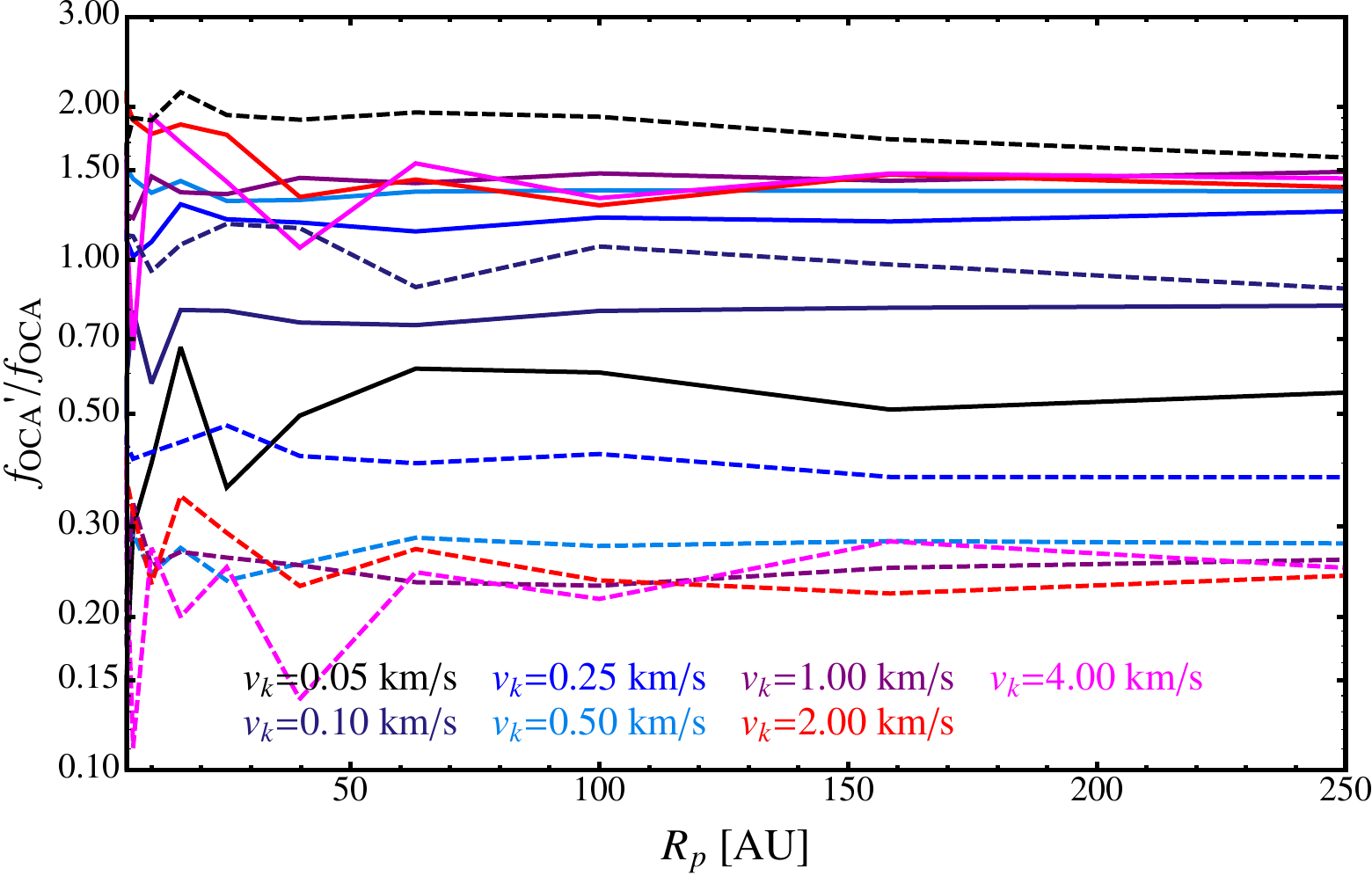}
\caption{Fraction of the mass of the OCA, $f_{\rm OCA}'={\rm d}N/{\rm dln}R_{\rm p}$, deposited on orbits with pericenter radius $R_{\rm p}$ following the natal white dwarf kick, calculated for two (non-fiducial) assumptions regarding the radial distribution of the OCA mass.  The result is expressed as a fraction of the fiducial distribution $f_{\rm OCA}$ (Fig.~\ref{periBoth}) and shown with the same color coding.  Solid lines correspond to a more centrally concentrated OCA ($\gamma=4.0$), while the dashed lines represent a shallower OCA profile ($\gamma=2.5$), than the fiducial case, $\gamma = 3.5$.  The limited resolution of the Monte Carlo sample is apparent as artificial noise at low values of $R_{\rm p}$.}
\label{periExplo}
\end{figure}

Figure \ref{periBoth} shows the fraction of the OCA mass, $f_{\rm OCA}={\rm d}N/{\rm dln}R_{\rm p}$, deposited per log pericenter radius $R_{\rm p}$ following the kick, calculated for a range of kick velocities, assuming our fiducial model for the initial OCA properties.  For the low pericenter radii of relevance, the distribution of comets is found to be roughly flat per unit pericenter, independent of kick velocity.  However, there is an important dichotomy: when $v_{\rm k} \lesssim 1~{\rm km~s}^{-1}$, almost all of the low-$R_{\rm p}$ orbits are elliptical, but when $v_{\rm k} \gtrsim 2~{\rm km~s}^{-1}$, the vast majority are hyperbolic.  This has potentially important implications for debris evolution, although we argue in \S 3.2 that the large thermal spread in debris energy reduces the importance of this distinction.

The ``sweet spot'' for impulsive velocity perturbations occurs for $v_{\rm k} \approx 0.5~{\rm km~s}^{-1}$.  Kick velocities significantly below this value are too weak to maximize the fraction of OCA objects put onto nearly radial orbits.  For kick velocities significantly above this, almost all OCA objects are on hyperbolic orbits, and the velocity space ``loss cone'' for nearly-radial orbits shrinks in size as $v_{\rm k}$ increases further.  Intuitively, this sweet spot corresponds to the orbital velocity near the inner edge of the Oort cloud (or at the transition radius between impulsive and semi-adiabatic kicks, whichever is larger).

Figure \ref{periExplo} shows the post-kick OCA mass distribution calculated for steeper ($\gamma=4.0$) and shallower ($\gamma=2.5$) density profiles than the fiducial case (Fig.~\ref{periBoth}).  By concentrating more comets at the inner edge, the steeper distribution gives an enhanced fraction $f_{\rm OCA}'$ ($\approx 50\%$ greater than the fiducial case) of OCA objects perturbed onto orbits with a given pericenter $R_{\rm p} \ll R_{\rm in}$.  Likewise, the shallower power law index $\gamma=2.5$ results in a $\approx 50-80\%$ decrease in $f_{\rm OCA}'$ for $v_{\rm k} \gtrsim 0.1$ km s$^{-1}$, with the decrease becoming noticeably larger for greater kick velocities.  A different value of $\gamma$ also shifts the sweet spot for $v_{\rm k}$; the steeper profile increases the kick velocity needed to maximize $f_{\rm OCA}'$, while the shallower profile reduces it.

An important consideration in these calculations is the net angular momentum of the comets placed onto nearly radial orbits.  If (as we will argue in the following sections) a large fraction of these comets sublimate and their orbital energy is dissipated, then their gaseous debris will eventually circularize at a radius determined by both the net post-kick angular momentum, $\vec{J}_{\rm tot}$, and the timescale for different gaseous debris streams to redistribute their own angular momentum in shocks.  A stable, circular disk will form if $|\vec{J}_{\rm tot}|>0$, as is possible only if the initial OCA possesses a net angular momentum that is not aligned with the direction of the WD kick.  Even if $|\vec{J}_{\rm tot}|=0$, short-lived gas disks may be able to form with very small radii due to the finite number of evaporating comets, and their resulting Poissonian excess in angular momentum.  Multiple eccentric gas disks may be able to form if shocks are inefficient at redistributing angular momentum, but the dissipative interaction of these with each other will likely combine them into a single disk or inflow quickly.  If $\vec{J}_{\rm tot}=0$ due to the underlying symmetry of the system, gaseous cometary debris will likely accrete very quickly, either falling directly onto the WD surface or forming a very compact accretion disk.  

Observations suggest that the net angular momentum of the Solar Oort Cloud is very small, as determined by the fraction $0.501 \pm 0.051$ of long period comets that reside on prograde orbits with respect to the ecliptic \citep{WieTre99}.  This observation is in tension with theoretical models of the formation of the Oort Cloud, which generally predict an excess of comets on retrograde orbits due to the preferential ejection of prograde comets as a result of their longer gravitational encounters with massive planets \citep{WieTre99}.  The numerical simulations of \citet{Brasse+06} have found some net angular momentum to the inner Oort Cloud, albeit with error bars comparable in size to the angular momentum in question.  More recent analysis of observations found no significant net angular momentum in observed Oort comets and suggested that past claims of a prograde bias have been due to selection effects \citep{WanBra14}.

We have replicated our fiducial scenario using an OCA with varying degrees of net rotation, by placing a fraction $f_{\rm rot}$ of comets on orbits with a preferred angular momentum direction\footnote{More specifically, we draw a fraction $f_{\rm rot}$ of our comets from the same distributions of orbital elements, but remove (and redraw) the $50\%$ of that subsample which is retrograde with respect to an arbitrary direction.}.  We find that the resulting distribution of comets on nearly-radial orbits preserves that specific angular momentum, such that the set of all comets with pericenters near $R_{\rm p}$ possess an average specific angular momentum $\approx f_{\rm rot}\sqrt{2GM_{\rm WD}R_{\rm p}}$.

\subsection{Non-Impulsive Mass loss}
\label{sec:nonImpulsive}

If WD natal kicks are delivered during an AGB superwind phase of duration $\approx 3\times 10^4~{\rm yr}$, then we can compare this timescale to an object's orbital time to estimate the validity of the impulsive approximation.  As in \citet{VerWya12}, we define the adiabaticity parameter $\Psi=t_{\rm orb}/t_{\rm SW}$.  The impulsive limit described in the prior section formally corresponds to $\Psi \gg 1$, but for an object orbiting a central mass of $M=0.6M_{\odot}$ at the inner edge of an OCA ($a=2000~{\rm AU}$), $\Psi=3.8$ and the star's mass loss will appear only moderately impulsive.  To determine the error this introduces into our estimates of $f_{\rm OCA}$, we have conducted simple two-body integrations for test particle orbits around central stars being kicked in a given direction over an adjustable timescale $t_{\rm k}$.  

In general, we find that a narrow window of initial true anomalies $f$ can be excited to near-parabolic eccentricities by central kicks.  The precise location of this band of $f$ does not asymptote to the impulsive limit until $\Psi \gtrsim 10$, but the width of the band does not change dramatically as long as it exists.  Indeed, in the trans-adiabatic regime, the bias is towards a widening.  However, once a kick becomes sufficiently non-impulsive, the initial orbital phase $f$ does not matter and comets with all possible initial phases reach the same final eccentricity $e'$.  

\begin{figure}
\includegraphics[width=85mm]{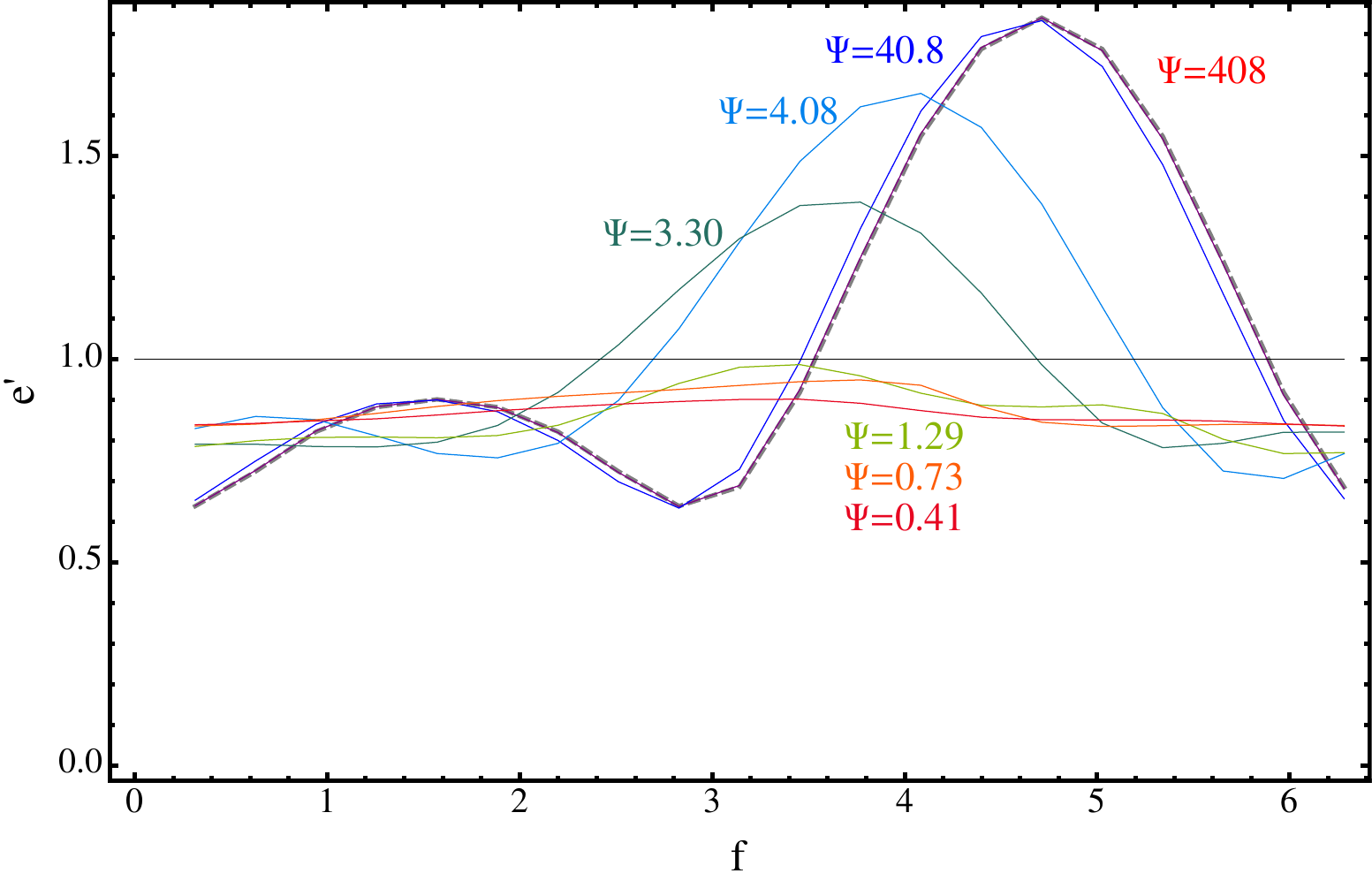}
\caption{Final eccentricity $e'$ as a function of initial orbital phase $f$, for test particles on $a=1000~{\rm AU}$, $e=0$ orbits around a central body receiving a kick $v_{\rm k}=0.5~{\rm km~s}^{-1}$.  The different colored lines represent different kick timescales $t_{\rm k}$; the red, blue, cyan, green, yellow, orange, and red lines indicate $t_{\rm k}/{\rm yr}$ values of $10^2$, $10^3$, $10^4$, $10^{4.25}$, $10^{4.5}$, $10^{4.75}$, and $10^5$, respectively.  The corresponding adiabaticity parameter values are labelled in the plot.  The thick, dashed gray line (overlapping with the $\Psi=408$ curve) represents the analytic prediction for the fully impulsive limit, as can be calculated using Appendix B.  All values of $\Psi>1.29$ produce eccentricity excitation to near-parabolic values, and the $\Psi=1.29$ case barely misses the window.}
\label{fig:transadiabatic}
\end{figure}

We plot the dependence of $e'$ on $f$ for a wide range of $\Psi$ in Fig. \ref{fig:transadiabatic}, in the special case where initial eccentricity $e=0$ and the kick is within the orbital plane.  We find that an impulsively excited band of $f$ will produce $e'$ arbitrarily close to 1 so long as $\Psi > \Psi_{c}\approx 1.5$.  Test integrations with different initial eccentricities and mutual inclinations (between the kick and the orbit) change the critical adiabaticity $\Psi_{\rm c}$ by factors of a few.  For the innermost edge of our fiducial model after mass loss ($a=2000~{\rm AU}$), and orbits with inclinations $i=\pi/4$, we find that the minimum pericenter qualitatively consistent with the impulsive approximation (i.e. capable of producing radial orbits) is $\approx 500~{\rm AU}$.

We note that a much more thorough treatment of anisotropic mass loss in non-impulsive regimes has been recently given by \citet{Veras+13}.  The analytic formalism presented in that paper would likely be necessary to investigate OCAs with $R_{\rm in} \lesssim 1000~{\rm AU}$, but that is beyond the scope of the present work.

\section{Destruction of Deeply Plunging Comets}

OCA objects perturbed onto orbits with sufficiently low angular momentum are vulnerable to tidal disruption and sublimation.  Both destructive processes depend on the mass $M_{\rm c}$ and radius $R_{\rm c}$ of the comet, with smaller objects being generally more vulnerable to sublimation (and larger ones to tidal disruption).  Only limited information is available about the size distribution of objects in the Solar Oort Cloud as compared to the Kuiper Belt, so the latter is used to guide our work.  The observed KBO size distribution is well modeled as an unfinished collisional cascade acting on a primordial distribution of KBOs \citep{Schlic+13}.  The primordial distribution, as set by coagulation with runaway growth among more massive bodies \citep{SchSar11}, is well approximated as 
\begin{equation}
N_{\rm pri}(R_{\rm c}) \propto R_{\rm c}^{-4} \label{primordialMF}
\end{equation}
above a certain radius $R_{\rm c} \sim 5~{\rm km}$ defining the transition to gravitationally focused growth.\footnote{The size distribution possesses an additional bump below this gravitationally-focused radius, which we conservatively ignore.}

After $4.5~{\rm Gyr}$ of collisional evolution, many solar KBOs have been eroded.  Theoretical work by \citet{Schlic+13} finds that the current distribution can be parameterized as a piecewise function
\begin{equation}
N_{\rm SFT}(R_{\rm c}) \propto 
\begin{cases}
R_{\rm c}^{-3.7}, & R_{\rm c}\le 0.1~{\rm km}\\  \label{SFTMF}
R_{\rm c}^{-2.5}, & 0.1~{\rm km} < R_{\rm c} \le 2~{\rm km} \\
R_{\rm c}^{-5.8}, & 2~{\rm km} < R_{\rm c} \le 10~{\rm km} \\
R_{\rm c}^{-2}, & 10~{\rm km} < R_{\rm c} \le 30~{\rm km} \\
R_{\rm c}^{-4}, & R_{\rm c} > 30~{\rm km} \\
\end{cases}
\end{equation}
The Solar Oort Cloud (and, presumably, the extrasolar OCAs of interest) were scattered out of a planetesimal disk at early times in the history of the system.  The primordial $N_{\rm pri}(R_{\rm c})$ and the collisionally processed $N_{\rm SFT}(R_{\rm c})$ distributions can thus be taken to bracket the true OCA distribution.  

The minimum of the size distribution on the main sequence will be $R_{\rm c, min}\approx 200 {\rm \mu m}$, as set by orbital decay due to PR drag over the OCA formation time $\sim 1$ Gyr (assuming a solar type star and distance $50~{\rm AU}$; see \S 4.3).  This may not be a firm minimum due to collisional replenishment of small grains during OCA formation \citep{BonWya10}; however, post-main sequence evolution will eliminate many of the smaller OCA particles.  In particular, radiation pressure blowout on the AGB branch and drag from dust-driven winds will (at $\sim 1000~{\rm AU}$ distances) both remove particles $\lesssim 1~{\rm cm}$ in size \citep{BonWya10}.  We therefore take $R_{\rm c, min}= 1~ {\rm  cm}$, and set the maximum size to $R_{\rm c, max} = 1000 ~{\rm km}$, roughly the upper limit of observed KBOs in the Solar System.  More details of OCA formation are discussed in Appendix A.

For the remainder of this paper, all our comets are treated as objects with constant density $\rho_{\rm c}=0.6~{\rm g~cm}^{-3}$, to match observations of Solar System comets \citep{DavGut04, AHearn+05, DavGut06}.  This is significantly lower than the density of their constituent solid-state silicates and ices, $\rho_{\rm d}=3~{\rm g~cm}^{-3}$.  This discrepancy arises because comets are generally loosely bound rubble piles with high internal porosity.

\subsection{Tidal Disruption}

A self-gravitating body of mass $M_{\rm c}$ and radius $R_{\rm c}$ is tidally disrupted by a white dwarf if the pull of tidal forces exceeds the object's internal gravity, i.e. at orbital radii $R$ less than the tidal radius
\begin{equation}
R_{\rm t}^{\rm SG}=R_{\rm c} \left( \frac{M_{\rm WD}}{M_{\rm c}} \right)^{1/3},
\label{tidal}
\end{equation}
where $M_{\rm WD}$ is the WD mass.  This critical radius can be thought of as the distance interior to which the orbital frequency $\sqrt{GM_{\rm WD}/R^3}$ exceeds the characteristic internal frequency $\sqrt{GM_{\rm c}/R_{\rm c}^3}$.  

Equation (\ref{tidal}) cannot be applied to most comets, which are smaller objects held together by solid state forces that instead respond to tidal perturbations with an internal frequency of $c_{\rm s}/R_{\rm c}$, where $c_{\rm s}$ is the (solid state) sound speed of the comet.  This results in a different tidal radius for objects dominated by solid state forces, given by
\begin{equation}
R_{\rm t}^{\rm SS}=G^{1/3} M_{\rm WD}^{1/3} R_{\rm c}^{2/3}c_{\rm s}^{-2/3}.
\end{equation}
The transition between the self-gravitating and solid state regimes occurs for objects roughly the size of Ceres, with masses $M_{\rm c} \sim 10^{24}~{\rm g}$ and radii $R_{\rm c} \sim 5\times 10^5~{\rm m}$.  In general, $R_{\rm t}^{\rm SS} <  R_{\rm t}^{\rm SG}$ if $R_{\rm t}^{\rm SG}$ is calculated using the mass and radius of a solid state-dominated body.  The exact tidal disruption radius for bodies bound by solid state forces is a complex function of rotation; material ductility, which determines how much deformation is necessary to produce catastrophic splitting; and material density, which can vary from our fiducial $0.6~{\rm g~cm}^{-3}$ depending on porosity of the rubble pile.  

Much more precise treatments of tidal disruption in the solid state limit exist \citep[for example]{Davids99}, but these do not change the fact that the tidal disruption radius is extremely small ($R_{\rm t}^{\rm SG} \sim 3\times 10^{-3}~{\rm AU}$, for the Ceres example).  In practice, a much greater number of comets are destroyed through sublimation, which we focus on hereafter.

\subsection{Evaporation}
\label{sec:evap}

A comet orbiting the white dwarf spends a time
\begin{equation}
t_{\rm p} = 2\pi \left( \frac{R_{\rm p}^3}{GM_{\rm WD}} \right)^{1/2} = 1.3R_{\rm p,AU}^{3/2}M_{0.6}^{-1/2}\,\,{\rm yr}
\end{equation}
near a pericenter radius $R_{\rm p}$, where $M_{0.6} = M_{\rm WD}/0.6M_{\odot}$ and $R_{\rm p, AU}=R_{\rm p}/{\rm AU}$.  Its equilibrium temperature is given by
\begin{equation}
T_{\rm eq}=890~{\rm K}~L_2^{1/4}R_{\rm p, AU}^{-1/2}~,
\end{equation}
where $L_{\rm WD}=100L_2~L_{\odot}$ is the WD luminosity.  The luminosity of the young WD decays as approximately a power-law with age $t_{\rm WD}$,
\begin{equation}
L_{\rm WD}=L_0 \left( \frac{t_{\rm WD}}{10^5~{\rm yr}} \right)^{-\lambda}.
\label{eq:LWD}
\end{equation}
Typical values are $L_0=10^2 L_{\odot}$, $\lambda=1.25$  (\citealt{Althau+09}).  For simplicity, the remainder of this paper implements a time-independent model for comet evaporation and debris dynamics, where $L_{\rm WD}$ is a free parameter.  Although a full time-dependent model would give more precise results, the fact that, generally, $\lambda > 1$ means that we can capture the basic picture in a time-independent way; later calculations in this paper establish that both evaporation of comets and orbital evolution of solid debris (due to radiation forces) are dominated by the first pericenter passage.

Ice and silicates sublimate at temperatures of $T_{\rm ice}\approx 170~{\rm K}$ and $T_{\rm rock} \sim 1500{\rm K}$, respectively, i.e. at radii interior to
\begin{equation}
R_{\rm ice} = 30L_2^{1/2}~{\rm AU}, \label{rIce}
\end{equation}
\begin{equation}
R_{\rm rock} = 0.3L_2^{1/2}~{\rm AU}.
\label{rock}
\end{equation}

The time required at an orbital distance of $R < R_{\rm ice}(R_{\rm rock})$ for the ice (rock) of a comet to completely sublimate is approximately given by
\begin{equation}
t_{\rm ev} = \frac{16\pi}{3} \frac{R_{\rm c} Q_{\rm C, V} \rho_{\rm c} R^2}{L_{\rm WD}} = 1.8 \times 10^7~{\rm s}~ R_{\rm c, km} L_2^{-1} \left( \frac{R}{\rm AU} \right)^2,
\end{equation}
where $R_{\rm c}=10^5~{\rm cm}~R_{\rm c, km}$ is the comet radius and $Q_{\rm C, V} \approx 3\times 10^{10}~{\rm erg}~{\rm g}^{-1}$ is the latent heat of transformation, which is similar for both ice and silicates.  

For the highly eccentric orbits under consideration, energy deposition near pericenter dominates the sublimation process because the heat deposited in the comet at an orbital radius $R$ is $Q \propto t(R)/R^2 \propto R^{-1/2}$.  By equating $t_{\rm ev} = t_{\rm p}$, we find that comets fully sublimate on their first pericenter passage interior to a `sublimation radius' given by
\begin{equation}
R_{\rm ev} = 5.4~{\rm AU}~L_2^2R_{\rm c, km}^{-2}. 
\label{evapRadius}
\end{equation}
Realistic comets consist of a mixture of silicates and volatiles, but the sublimation of cometary ices is likely a sufficient criterion for their destruction.  Comets will therefore sublimate efficiently when $R_{\rm p}< \min (R_{\rm ev}, R_{\rm ice} )$.  

Comets with pericenters obeying $R_{\rm ev} < R_{\rm p} < R_{\rm ice}$ can also sublimate, but less efficiently and over many pericenter passages.  Appendix C shows that this results in an enhancement of at most a factor of a few to the total mass lost from the largest comets; we include this modest enhancement to $R_{\rm ev}$ (assuming $\lambda=1.25$, $R_{\rm ev}$ increases by a factor $\approx 1.5$)\footnote{Multiple passages occur only for comets on bound orbits and hence are therefore irrelevant for kick velocities $v_{\rm k} \gtrsim 2~{\rm km~s}^{-1}$.}.  

\begin{figure}
\includegraphics[width=85mm]{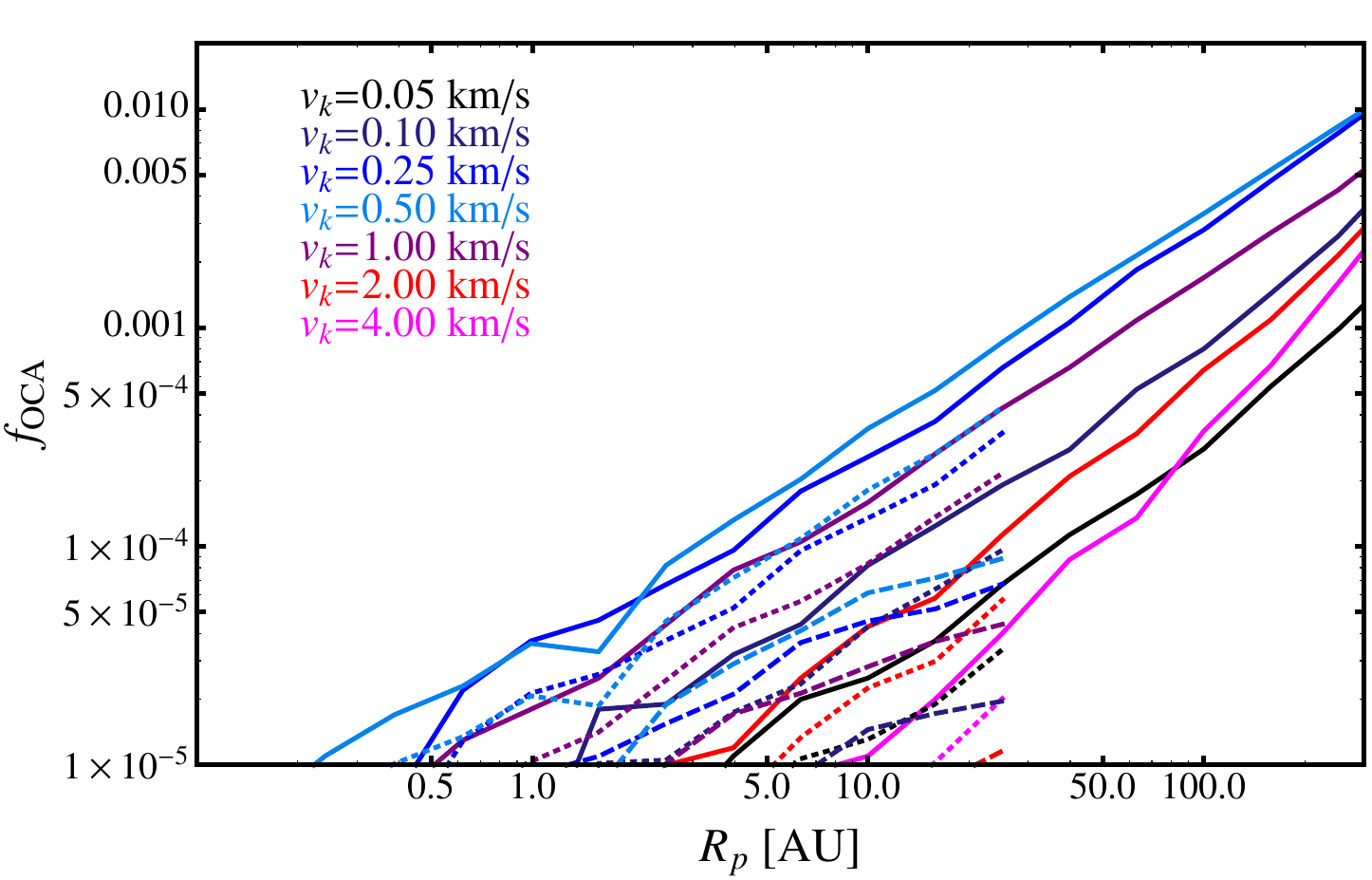}
\caption{Distribution of mass {\it sublimated} from comets as a function of the pericenter radius $R_{\rm p}$ in comparison to the total distribution of kicked-in comets $f_{\rm OCA}={\rm d}N/{\rm dln}R_{\rm p}$ ({\it solid lines}; same color labeling as Fig.~\ref{periBoth}).  The fraction of comets that sublimate is calculated according to criteria given in Eq. (\ref{evapRadius}) taking into account the size distribution of comets.  Dotted lines are calculated assuming the optimistic `primordial' mass function Eq.~(\ref{primordialMF}), while the dashed lines show the pessimistic, `collisionally processed' mass function Eq.~\ref{SFTMF}).  Both dotted and dashed lines truncate at the ice line for $L_{\rm WD}=10^2L_{\odot}$; if $R_{\rm p}>R_{\rm ice}$, no sublimation is possible.}
\label{massFunction}
\end{figure}

Figure \ref{massFunction} shows the mass of the kicked-in OCA that actually sublimates in a given range of pericenter radii, calculated according to the criterion $R_{\rm c}<R_{\rm ev}$ for the two mass functions (Eqs.~\ref{primordialMF},\ref{SFTMF}) which bracket our ignorance of the amount of collision processing experienced by the OCA.  The less collisionally processed distribution sees $\sim 50\%$ of the kicked-in mass at a given pericenter sublimate, but for the top-heavy mass distribution of Eq. (\ref{SFTMF}), only $\sim 10\%$ sublimates.  

The debris from sublimating comets can only produce observable consequences, such as forming a gaseous or dusty disk, if it remains bound to the WD.  However, Figure \ref{periBoth} shows that the vast majorities of comets are placed onto hyperbolic trajectories for kick velocities $v_{\rm k} \gtrsim 1~{\rm km~s}^{-1}$.  As we now discuss, the sublimation processes imparts a spread in the specific orbital binding energy $\Delta \epsilon$ of the sublimated gaseous/dusty debris which can in some cases exceed the center of mass specific orbital energy $| \epsilon_0|$.  In this case a significant fraction, $\sim 1/2$, of the debris (that imparted with negative energy) will remain bound to the WD, even if the initial center of mass orbit of the comet was not.  

The outgassing volatiles from a sublimating comet will possess a specific energy (in the rest frame of the comet) similar to the thermal energy associated with its sublimation temperature.  Upon decompression they will thus obtain a velocity similar to the thermal velocity of the gaseous component $v_{\rm th}=\sqrt{3k_{\rm B}T_{\rm ice}/m_{\rm ice}}$, where $k_{\rm B}$ is the Boltzmann constant and $m_{\rm ice} \approx 18m_{\rm p}$ is the mass of the sublimating ice\footnote{More detailed models for comet evaporation show that volatiles reach a terminal ejection velocity of $v_{\rm ej, g} = \sqrt{\frac{\gamma_{\rm a}+1}{\gamma_{\rm a}-1} \frac{\gamma_{\rm a}k_{\rm B}T}{m_{\rm ice}}}$,
where the ratio of specific heats is $\gamma_{\rm a} \approx 4/3$ \citep{CriRod97}.  This is in fairly good agreement with observations of comets outgassing within the Solar System \citep{CriRod99}.}.  However, the ejection velocity of solid debris may differ from that of the gas \footnote{The ejection velocity for dust, $v_{\rm ej, d}$, is generally lower, and is quite model-dependent.  A review of different dust ejection models is provided in \citet{Ryabov13}.  For grain sizes $b \sim 100~{\rm \mu m}$, these generally predict $5\times 10^{-2}~{\rm km~s}^{-1} \lesssim v_{\rm ej, d} \lesssim10^{-1}~{\rm km~s}^{-1}$ (as we shall see in \S \ref{DD:ev}, this is the relevant grain size for our purposes).  Dust ejection velocities are larger for higher rates of outgassing, i.e. smaller pericenters.  Therefore, for large kick velocities ($v_{\rm k} \gtrsim 1~{\rm km~s}^{-1}$), it is possible that only a deeply plunging subpopulation will have bound solid debris.}.  For any cometary constituent ejected at a speed $v_{\rm ej}$, the resulting spread in its specific orbital energy will be
\begin{equation}
\Delta\epsilon \sim\max\left( \frac{1}{2}v_{\rm ej}^2, v_{\rm ej}V_{\rm p}\right),
\end{equation}
with $V_{\rm p}=\sqrt{2GM_{\rm WD}/R_{\rm p}}$ the orbital velocity at pericenter (where most sublimation occurs) for a highly eccentric orbit.  Unless $R_{\rm p}>10^4~{\rm AU}$ (in which case the comet would not sublimate in the first place), $v_{\rm ej}V_{\rm p}$ is the greater of these two quantities.  

The effective ejection velocity may also be set by the rocket effect, as volatiles outgassing from one side of a comet or comet fragment push the host body in the opposite direction.  Observations of the fragmentation of Comet 73P/Schwassmann-Wachmann 3B suggest that this reaction acceleration can be substantial, $\sim 10^{-3}$ of the solar gravitational acceleration at the time of the comet's breakup at $\approx 2~{\rm AU}$ \citep{Ishigu+09}.  If this acceleration were sustained for the pericenter passage of an exocomet through a newborn WD's ice line at $R_{\rm ice}=30~{\rm AU}$, it would produce an effective ejection velocity $v_{\rm ej, d} \approx 1~{\rm km~s}^{-1}$, enough to retain $\sim 1/2$ of the solid debris even for large $v_{\rm k}$.  In general,
\begin{align}
\frac{\Delta\epsilon}{{\rm erg}\,\,{\rm g}^{-1}} \sim 
\begin{cases} 2.0 \times 10^{11}M_{0.6}^{1/2}R_{\rm p, AU}^{-1/2} \left( \frac{T}{T_{\rm ice}} \right)^{1/2}, ~{\rm gas} \\ 
2.3 \times 10^{11}M_{0.6}^{1/2}R_{\rm p, AU}^{-1/2} \left( \frac{v_{\rm ej, d}}{1~{\rm km~s}^{-1}} \right), {\rm dust} . 
\end{cases}
\end{align}
This spread in energy exceeds the center-of-mass energy of the (hyperbolic) comet orbit $\sim | \epsilon_{\rm orb}| \lesssim v_{\rm k}^2/2$ interior to a critical pericenter radius
\begin{align}
\frac{R_{\rm hyper}}{\rm AU} \sim 
\begin{cases}
97~{\rm AU} ~M_{0.6} \left( \frac{T}{T_{\rm ice}} \right) \left( \frac{v_{\rm k}}{2~{\rm km~s}^{-1}} \right)^{-4}, ~{\rm gas} \\ \label{hyper}
133~{\rm AU} ~M_{0.6} \left( \frac{v_{\rm ej, d}}{1~{\rm km~s}^{-1}} \right)^2 \left( \frac{v_{\rm k}}{2~{\rm km~s}^{-1}} \right)^{-4}, ~{\rm dust}.
\end{cases}
\end{align}

\begin{figure}
\includegraphics[width=85mm]{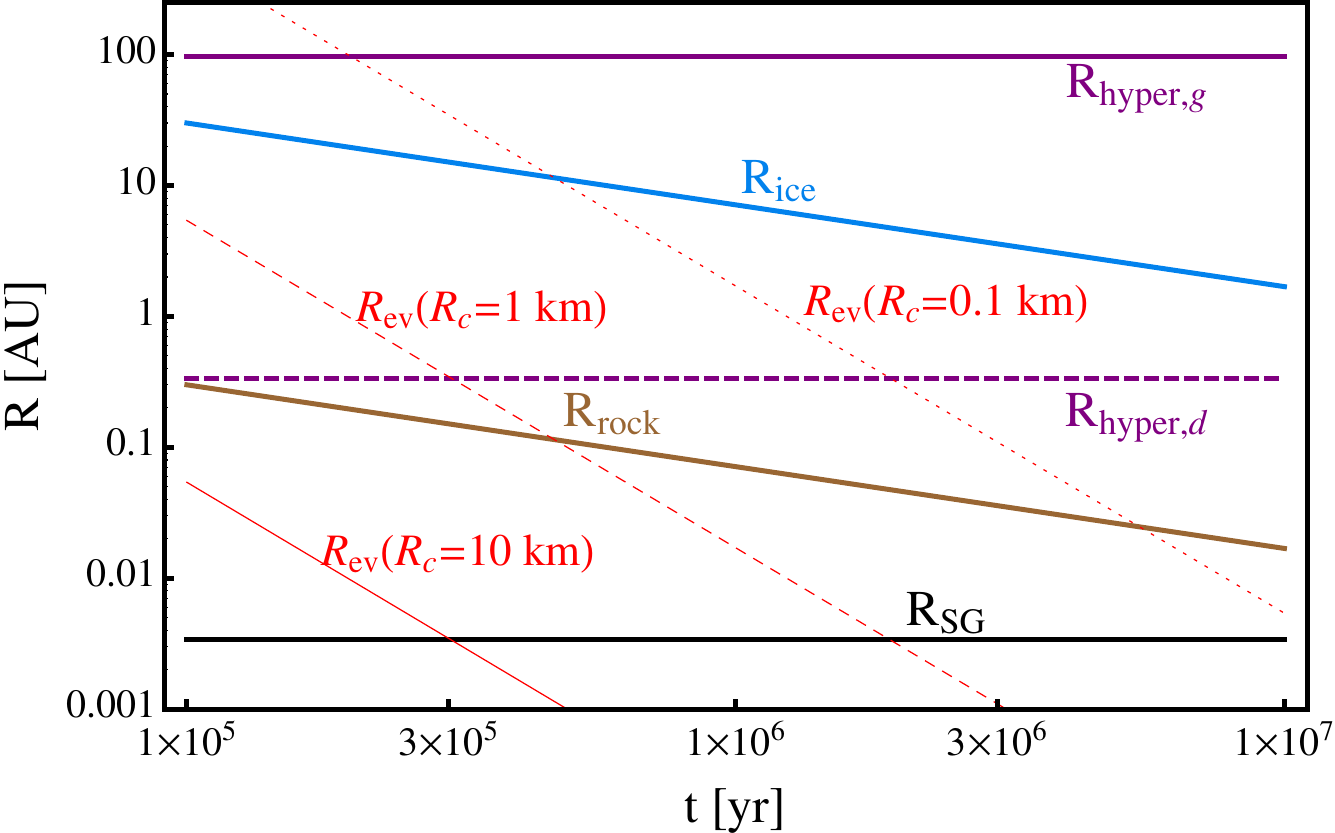}
\caption{Critical radii as a function of time (or, equivalently, the WD age).  Radii shown include the tidal radius for a self-gravitating body similar to Ceres, $R_{\rm t}^{\rm SG}$ Eq.~\ref{tidal}; {\it black solid}), sublimation radius for ice $R_{\rm ice}$ (Eq.~\ref{rIce}; {\it blue solid}), sublimation radius for silicates, $R_{\rm rock}$ (Eq.~\ref{rock}; {\it brown solid}), and the radius interior to which most debris from comets on initially hyperbolic orbits remains bound to the WD, $R_{\rm hyper}$ (Eq.~\ref{hyper}; {\it purple}).  Two different $R_{\rm hyper}$ curves are shown: one for sublimated volatiles ({\it purple solid}; this may also correspond to the dust hyperbolic radius if the rocket effect is strong), and one for ejected dust ({\it purple dashed}).  Red lines show the sublimation radii $R_{\rm ev}$ (Eq.~\ref{evapRadius}) for comets of size $R_{\rm c} = 10$ km ({\it solid}), $1$ km ({\it dashed}), and $0.1$ km ({\it dotted}), respectively.  We note that the sublimation radii have no physical meaning when located above their respective ice lines (for either water ice or silicates); for example, when $R_{\rm ev}>R_{\rm ice}$, portions of parameter space between these curves are where the comet absorbs enough (eventually re-radiated) energy at pericenter to fully sublimate, but cannot because its temperature is too low.}
\label{radii}
\end{figure}

For orbits with $R_{\rm p}<R_{\rm hyper}$, roughly half of the cometary debris will become unbound, while the other half will remain bound to the WD with a wide range of orbital eccentricities and semimajor axes.  For gas, this is generally the case when $v_{\rm k} \lesssim 3~{\rm km~s}^{-1}$.  If the rocket effect scatters a large fraction of solid debris, the same criterion holds; otherwise, most sublimating orbits will only satisfy $R_{\rm p}<R_{\rm hyper}$ if $v_{\rm k} \lesssim 1~{\rm km~s}^{-1}$.  Conversely, when $R_{\rm p}>R_{\rm hyper}$, all the debris will remain ``frozen in'' to orbits with roughly the same parameters as the center of mass trajectory, which are bound (unbound) for $v_{\rm k} \lesssim 1~{\rm km~s}^{-1}$ ($v_{\rm k} \gtrsim 1~{\rm km~s}^{-1}$).  Because $R_{\rm hyper} > R_{\rm ice}$ is generally satisfied for $v_{\rm k} \lesssim 3~{\rm km~s}^{-1}$, we conclude that to first order $50\%$ of all sublimated OCA gas will remain bound to the WD for these kick velocities\footnote{$50\%$ of solid debris may remain bound given this same criterion, but this outcome may require the slightly stricter criterion that  $v_{\rm k} \lesssim 1~{\rm km~s}^{-1}$, depending on the importance of the rocket effect in disintegrating comets.}.  When $v_{\rm k} \gtrsim 3~{\rm km~s}^{-1}$, a large majority of sublimated debris will instead exit the system, producing much weaker observational signatures.  For the remainder of this paper, we assume $50\%$ of debris remains bound, but we note here that if the rocket effect is unimportant, most solid debris will exit the system for $v_{\rm k} \gtrsim 1~{\rm km~s}^{-1}$


Under the assumption that the distribution of specific orbital energy of the sublimating debris is uniform\citep{Rees88}, the resulting semi-major axis distribution is given by
\begin{equation}
\frac{{\rm d}N}{{\rm d}a} = \frac{a_{\rm min}}{2a^2}, \label{aDist}
\end{equation}
where the semi-major axis of the most tightly bound debris is given by
\begin{equation}
a_{\rm min}=\frac{GM_{\rm WD}}{2\Delta \epsilon}=12.7~{\rm AU} ~M_{0.6}^{1/2}R_{\rm p, AU}^{1/2}, \left(\frac{T}{T_{\rm ice}}\right)^{-1/2}. \label{semimajorDist}
\end{equation}

The many critical radii discussed in this subsection are plotted as functions of time in Fig. \ref{radii}.  As the WD ages and cools, the sublimation radii for comets of fixed size contract quickly, while the ice lines move inward more slowly.

\subsection{Cometary Debris}
\label{sec:debris}

Once one or more comets have sublimated, multiple streams of gas and solid debris will orbit the WD on highly eccentric trajectories.  We parametrize the relative composition of these two constituents with $f_{\rm gas}$, the volatile mass fraction of the comet prior to sublimation.  We also account for the hydrogen mass fraction of the sublimated gas, $f_{\rm H}$.  We hereafter make the simplifying assumptions that $f_{\rm gas}=0.5$ and that water ice is the dominant volatile, implying $f_{\rm H}=f_{\rm gas}/8=0.06$.

Solid state debris left over from process of cometary sublimation possesses an initial distribution $n_{\rm ss}(b)$ of particle sizes $b$.  Studies of solar cometary sublimation typically model this distribution as either a single power law \citep{Dohnan69}, 
\begin{equation}
n_{\rm ss}(b) = n_0 \left( \frac{b_{\rm min}}{b} \right)^{L}, \label{dustPL}
\end{equation}
or as the slightly more complex ``Hanner law'' distribution \citep{Hanner83}
\begin{equation}
n_{\rm ss}(b) = n_0 \left( 1 - \frac{b_{\rm min}}{b} \right)^M \left( \frac{b_{\rm min}}{b} \right)^N,
\end{equation}
where $b_{\rm min}$ is the minimum particle size and $n_0$ is a normalization constant.  Generally $3.7 < N < 4.2$ for the observed debris of cometary sublimation within the Solar System \citep{Harker+02}, although $L=3.5$ is often assumed in the single power-law case \citep{Donald+13}.  For simplicity we adopt equation (\ref{dustPL}) with $L=3.5$, although similar quantitative results are found using the Hanner distribution.  

Electron microscopy analysis of interplanetary dust grains indicates a lower size limit of $b_{\rm min} \sim 0.1~{\rm \mu m}$ \citep{Bradley94}, which we take as our fiducial value.  The maximum dust particle size, $b_{\rm max}$, is hard to measure remotely due to its generally subdominant contribution to reradiated sunlight from sublimated comets.  However, NASA's Deep Impact mission to the comet 9P/Tempel 1 excavated a large crater on the comet's surface, enabling a rough calculation of $b_{\rm max}$ by comparing the ejecta mass estimates from volumetric considerations to those from IR emission \citep{Kupper+05}.  Although this calculation is subject to large uncertainties, the most recent analysis of Tempel 1 results indicates that $1~{\rm \mu m} \lesssim b_{\rm max} \lesssim 100~{\rm \mu m}$ \citep{Gicque+12}.  Subsequent Deep Impact observations of the coma of comet 103P/Hartley indicate the presence of much larger particles, up to $\sim 1~{\rm m}$ in size.  However, the inferred power law slope of these ``boulders'' is so steep that they would contribute negligibly to both the mass and surface area budget of ejected dust.  In addition, density arguments favor an icy composition for the boulders, indicating that they would not be present for fully destructive sublimation events \citep{Kelley+13}.  As we will show in \S \ref{DD:ev}, only particles of size $b \gtrsim 50~{\mu m}$ can survive around young luminous WDs, motivating us to take $b_{\rm max}=200~{\rm \mu m}$ as fiducial.  If such large dust grains are not in fact present, then both the hypothesis of this paper (OCA origins) and the KBO model in \citet{Su+07} would have serious difficulty in explaining the {\it Spitzer} observations of extended dust halos around young WDs.

Under the above assumptions, the surface area per unit mass $\Upsilon$ of cometary dust is given by
\begin{equation}
\Upsilon=\frac{3}{\rho_{\rm d}} \frac{4-L}{3-L} \frac{\tilde{b}_{\rm max}^{3-L}-\tilde{b}_{\rm min}^{3-L}}{b_{\rm max}^{4-L}-b_{\rm min}^{4-L}},
\label{eq:upsilon}
\end{equation}
where $\tilde{b}_{\rm min}$ and $\tilde{b}_{\rm max}$ are the minimum and maximum grain size following post-evaporative processing (\S \ref{DD:ev}).  Here we have used our fiducial density for dust grains, $\rho_{\rm d}=3~{\rm g}~{\rm cm}^{-3}$.

\section{White Dwarf Natal Disks}

\subsection{Gaseous Disks: Evolution}
Assuming that a fraction $f_{\rm ev} \sim f_{\rm OCA} \sim 10^{-4}-10^{-3}$ of the total mass $M_{\rm OCA} \sim 10M_{\oplus} \sim 6\times 10^{28}$ g of the OCA sublimates (i.e. those comets with pericenter radii interior to the sublimation radius $R_{\rm ev}$; Fig.~\ref{massFunction}), then the resulting gas will circularize into a disk of mass $M_{\rm gas} = f_{\rm gas}f_{\rm ev}M_{\rm OCA} \sim 10^{25}$ g and characteristic radius $R_{\rm circ} \sim f_{\rm rot}^{2}R_{\rm ice} \sim 0.3(f_{\rm rot}/0.1)^{2}L_{2}^{1/2}$ AU, where $f_{\rm rot}$ is the rotation parameter of the original OCA (see \S \ref{sec:impulse}).  

The viscous accretion timescale at radius $r$ is given by
\begin{align}
t_{\rm visc} &\sim \frac{r^{2}}{\nu} \sim \frac{1}{\alpha \Omega_{\rm K}}\left(\frac{H}{r}\right)^{-2}  \\
&\sim 1.3\times 10^{3}\alpha_{-1} M_{0.6}^{1/2} \left(\frac{T_{\rm g}}{10^{4}{\rm K}}\right)^{-1}\left(\frac{r}{30~\rm AU}\right)^{1/2} {\rm yr},
\label{eq:tvisc}
\end{align}
where $\nu = \alpha c_{\rm s}H$ is the effective turbulent viscosity, $\Omega_{\rm K} \equiv (GM_{\rm WD}/r^{3})^{1/2}$, $\alpha = 10^{-1}\alpha_{-1}$ is the Shakura-Sunyaev viscosity parameter, $c_{\rm s} \simeq (k T/\bar{\mu})^{1/2}$ is the midplane sound speed, $\mu \approx 18 m_{\rm p}$ is the mean molecular weight of water, and $H = c_{\rm s}/\Omega_K$ is the vertical scale-height of the disk.  The disk temperature $T_{\rm g}$ is set by a competition between heating due to photo-ionization from the WD and cooling via line emission (e.g.~oxygen).  The inner parts of the disk are sufficiently dense to cool via optically-thick emission lines \citep[as in][]{Melis+10}, while the outer parts are optically thin and will instead cool via forbidden line emission, similar to an HII region \citep{OstFer06}.  In both cases, the equilibrium disk temperature is estimated to be $T_{\rm g} \sim 10^{4}$ K.

In the absence of other forms of mass loss, Eq. (\ref{eq:tvisc}) shows that a gaseous disk of radius $r \sim R_{\rm circ} \sim 0.3$ AU will accrete on a timescale $\sim 10^{2}$ years, much less than the age of the system, $\sim 10^{5}-10^{6}$ years.  Even the residual disk left over after viscous spreading at $r\sim 30~{\rm AU}$ will accrete quickly.  During this time the inflowing material will achieve a steady-state accretion rate $\dot{M} \sim M_{\rm gas}/t_{\rm fall} \sim 10^{12}$ g s$^{-1}$ ($t_{\rm fall} \sim 10^5~{\rm yr}$ is the fallback time for typical OCA comets post-kick) with a surface density profile at radii $r \lesssim R_{\rm circ}$ given by
\begin{eqnarray}
\Sigma_{\rm gas} &=& \frac{\dot{M}}{3\pi \nu}\label{eq:sigmagas}  \\
&\simeq& 2\times 10^{-7}\frac{\dot{M}}{10^{12}\,\rm g\,s^{-1}} M_{0.6}^{1/2}\alpha_{-1}^{-1}\left( \frac{r}{30~{\rm AU}} \right)^{-3/2}~{\rm g\,cm^{-2}}, \notag
\end{eqnarray}
where we have used Eq. (\ref{eq:tvisc}) assuming $T_g = 10^{4}$ K.

The disk midplane will be photo-ionized by ultraviolet radiation from the WD, as is justified because the radial optical depth of the disk $\tau \sim (\Sigma_{\rm g}/H)\kappa_{\rm es}r$ is less than unity at radii of interest, where $\kappa_{\rm es} \approx 0.2$ cm$^{2}$ g$^{-1}$ is the electron scattering opacity.  Heating from ionization drives a powerful outflow from the disk exterior to the characteristic radius $R_g \simeq GM_{\rm WD}/c_s^{2} \sim 30(T_{\rm g}/10^4~{\rm K})^{-1}$ AU \citep{Hollen+94}, with a mass loss rate $\dot{M}_w$ that is exponential in $r/R_{\rm g}$ (more detailed calculations find that $\dot{M}_w$ declines slowly with decreasing radius, until cutting off sharply at radii $r < R_{\rm g}'\approx 0.15R_{\rm g} \approx 5(T/10^4~{\rm K})^{-1}~{\rm AU}$; \citealt{Adams+04}).  The rate of mass loss due to photo-evaporation is sufficiently high that the outer disk ($r \gtrsim R_{\rm g}'$) will dissipate on a timescale much shorter than the age of the system.  Although $R_{\rm g}' \gg R_{\rm circ}$, viscous spreading will eventually bring some mass to radii vulnerable to photoevaporation.  However, the bulk of the gas that circularizes interior to $R_{\rm g}'$ will accrete onto the WD, guaranteeing that a minimum H fraction be available to pollute the WD atmosphere.

In this subsection, we have treated these gas disks in an approximate manner, with the largest uncertainty being the unknown value of $f_{\rm rot}$.  However, in the following two sections we shall see that even a very approximate treatment is adequate, because (i) even if a mass $M \ll M_{\rm gas}$ accretes onto the WD, there will still be observational consequences, and (ii) the effect of these disks on solid state debris is negligible.

\subsection{Gaseous Disks: Observational Implications}

The atmospheres of many WDs possess extremely low hydrogen abundances (e.g.~\citealt{Berger+11}); DB atmospheres with less than $10^{-16}M_{\odot}$ of hydrogen are not uncommon, and the most hydrogen depleted systems have upper limits below $10^{-18}M_{\odot}$.\footnote{DBA WDs have higher quantities of atmospheric hydrogen, but still generally less than $10^{-10}M_{\odot}$ \citep{Berger+11}.}  The accretion of a single moderately-sized comet would overproduce the observed hydrogen in many DB atmospheres, a fact which is puzzling in light of the comet accretion mechanisms described in previous sections.  Because hydrogen does not sediment away as is the case with metals, extremely hydrogen-deficient WD atmospheres provide tight constraints on the joint parameter space of WD kick velocities and OCAs.  

How much hydrogen will a natal kick deliver to the WD surface?  The total hydrogen mass, $M_{\rm H}$, accreted onto the WD via the transient gas disk discussed in the previous section is just that which will circularize interior to the photo-evaporation radius $R_{\rm g}' \approx 5$ AU.  Since $R_{\rm circ} \ll R_{\rm g}'$ generally, $M_{\rm H} \approx M_{\rm gas} f_{\rm H}$ where $f_{\rm H} \approx 0.06$ is the H fraction of typical OCA bodies.  For a characteristic value $M_{\rm gas}=10^{25}~{\rm g}$, we have $M_{\rm H} \sim 6 \times 10^{23}~{\rm g}$.  This expected level of pollution is many orders of magnitude higher than the observed H abundances in DB WD atmospheres, and at least comparable to (often higher than) H abundances in DBA atmospheres \citep{Berger+11}.  

\subsection{Debris Clouds: Evolution}
\label{DD:ev}
The fate of the solid debris is more complex.  A spherical solid of radius $b$ on a circular orbit of radius $r$ is dragged inwards via Poynting-Robertson (PR) drag on a timescale \citep{Rafiko11}:
\begin{equation}
t_{\rm PR} \underset{e = 0}\simeq \frac{4\rho_{\rm d} r^{2}c^{2}b}{L_{\rm WD}} \approx 2 \times 10^{5}L_{2}^{-1}\left(\frac{b}{\rm 100\mu m}\right)\left(\frac{r}{\rm 10~AU}\right)^{2}{\rm yr}
\label{eq:tPR}
\end{equation}
However, the PR timescale for highly eccentric orbits ($e \approx 1$) of the same pericenter radius,
\begin{align}
t_{\rm PR} \underset{e \approx 1}\approx  8\times 10^5 L_{2}^{-1}\left(\frac{b}{100\mu m}\right)\left( \frac{a_0}{10^3~{\rm AU}} \right )^{1/2} \left( \frac{R_{\rm p, 0}}{10~{\rm AU}} \right)^{3/2}\,{\rm yr},
\end{align}
can be significantly longer than in the circular case (Appendix D), where $a_0$ and $R_{\rm p, 0}$ are the initial semimajor axis and pericenter, respectively.

Since $L_{\rm WD} \propto t^{-1.25}$ for the cooling white dwarf (Eq.~\ref{eq:LWD}), the ratio $t_{\rm PR}/t$ is approximately independent of time.  Thus, by a time $t$, and outside of a radius $R_{\rm p, 0}$, PR drag will remove all particles with radii larger than
\begin{equation}
b_{\rm min}^{\rm PR}=13L_{2}\left(\frac{t}{10^5~{\rm yr}}\right)\left( \frac{R_{\rm p, 0}}{10~{\rm AU}} \right)^{-3/2} \left( \frac{a_0}{10^3~{\rm AU}} \right)^{-1/2} ~{\rm \mu m} .
\label{eq:bminPR}
\end{equation}
In addition to PR drag, particles can be removed by radiation blow-out if the WD luminosity is exceeds the Eddington luminosity for a particle cross section $\pi b^{2}$.  Radiation pressure thus sets its own minimum pebble size of 
\begin{equation}
b_{\rm min}^{\rm rad} = 32L_{2}M_{0.6}^{-1} ~{\rm \mu m} . \label{radBlowout}
\end{equation}

\begin{figure}
\includegraphics[width=85mm]{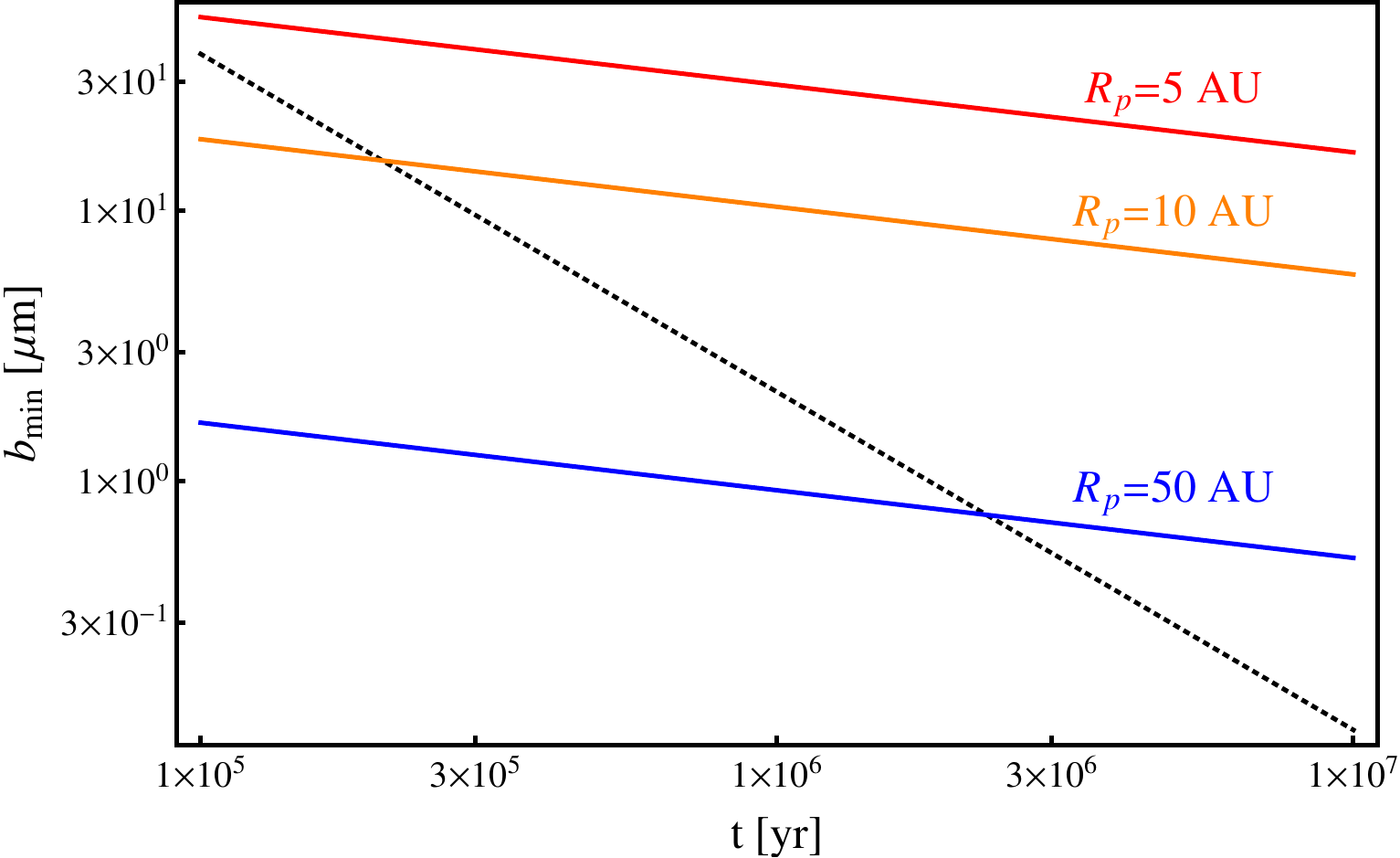}
\caption{Minimum grain size $b_{\rm min}$ that survives Poynting Robertson drag ({\it solid lines}; Eq.~\ref{eq:bminPR}) and radiation blow-out ({\it dotted black line}; Eq.~\ref{radBlowout}) as a function of the white dwarf age, calculated using the WD cooling evolution in Eq.~(\ref{eq:LWD}) for orbits with different pericenter radii: $R_{\rm p,0} = 5~{\rm AU}$ ({\it red}), $10~{\rm AU}$ ({\it orange}), and $50~{\rm AU}$ ({\it blue}), respectively.  All cases are calculated assuming a PR efficiency $Q_{\rm PR}=1$, solid density $\rho_{\rm d} = 3~{\rm g~cm}^{-3}$, WD luminosity normalization $L_0=10^2L_{\odot}$ (Eq.~\ref{eq:LWD}), and initial semi-major axis $a_0=500~{\rm AU}$, the latter characteristic of the debris following energy freeze-in during evaporation.}
\label{fig:grainSize}
\end{figure}

Figure \ref{fig:grainSize} shows the time evolution of the minimum grain size set by different processes at several characteristic radii.  In general, radiation pressure blowout is the limiting factor at early times and large radii, while PR drag plays a larger role close to the WD and at late times, as the central luminosity declines.  The characteristic minimum size is $b_{\rm min} \sim$ tens $\mu {\rm m}$ on timescales $t \sim $ few $10^{5}$ yr, which characterize the free fall time for most OCA objects on nearly parabolic orbits.\footnote{A generally weaker constraint we ignore here is the entrainment of small grains by the relative flow of the interstellar medium; this produces a minimum grain size $\approx 10~{\rm \mu m}$ \citep{HowRaf14}.}

We estimate the timescale for an eccentric collisional cascade using Eq. (16) of \citet{Wyatt+10}, which determines whether or not the size distribution remains fixed in time.  If we take characteristic values $b_{\rm max}=100~\mu$m, total dust mass $\sim 10^{-3}M_\oplus$, and semimajor axis $a=100~{\rm AU}$, we can estimate the collisional cascade timescale if we know the catastrophic disruption threshold $Q_{\rm D}^*$ and the dimensionless function $S(e)$.  We take $S(e)=1$ (for $1-e = 0.3$; see \citealt[Fig. 3]{Wyatt+10}), and extrapolate Fig. 3 of \citet{BenAsp99} to $\sim 100~\mu$m scales to estimate $Q_{\rm D}^* \sim 5\times 10^4~{\rm J~kg}^{-1}$ for rocky dust.  These parameters give a collisional cascade timescale $t_{\rm cc} \approx 5\times 10^6~{\rm yr}$, which is greater than the age of the systems we are concerned with.  We assume a constant particle size distribution for the remainder of this paper, but note that collisional grinding of the solid debris over longer timescales could produce a reservoir for eventual metal accretion onto the WD.  Small dust grains produced over timescales $t > t_{\rm cc}$ may eventually accrete due to PR drag.

The ultimate appearance of the solid debris will depend on whether the dusty grains produced by sublimation remain on eccentric orbits or are circularized into a disk-like configuration via drag on the (circular) gaseous disk ($\S\ref{sec:dustydisk}$).  For solids on {\it circular} orbits, the gas drag timescale is given by \citep{Rafiko04} 
\begin{equation}
t_{\rm drag}^{\rm circ} = \Omega_{\rm K}^{-1} \frac{\rho_{\rm d} b}{\Sigma_{\rm gas}}.
\end{equation}
in the Epstein limit.  The Epstein limit is appropriate because the particle size $b$ is much smaller than the gas mean free path $\lambda_{\rm mfp} =\bar{\mu}/(\rho_{\rm gas}\sigma_{\rm mol})$, where $\sigma_{\rm mol} \approx 10^{-15}~{\rm cm}^2$ is the molecular cross section, and $\rho_{\rm gas} = \Sigma_{\rm gas}/2H$ is the midplane density of the gaseous disk (see Eq.~\ref{eq:sigmagas}).  

As in the case of PR drag, the circular drag timescale of Eq. (\ref{eq:sigmagas}) is not applicable to highly eccentric orbits.  Because matter on an eccentric orbit resides at a given radius $r$ for a characteristic time $t \propto r^{3/2}$, and since the per-orbit drag dissipation of energy scales as $\Sigma_{\rm g} \propto r^{3/2}$ (Eq.~\ref{eq:sigmagas}), each logarithmic radius interval would contribute equally to the total gas drag if the gas were spherically distributed.  However, because the gas in fact resides in a disk, the total drag is dominated by those brief intervals when the solid passes through the disk midplane.  Assuming that such passages occur at orbital pericenter, a more general timescale for drag on eccentric debris can be derived: 
\begin{align}
t_{\rm drag} =& \Omega_{\rm K}^{-1} \frac{\rho_{\rm d} b}{\Sigma_{\rm gas}} \left( \frac{R_{\rm p}}{a} \right)^{-3/2} \nonumber \\
=& 3\times10^{8}~{\rm yr}~\left(\frac{\dot{M}}{10^{12}{\rm g\,s^{-1}}}\right)^{-1}M_{0.6}^{-1}\alpha_{-1} \\ 
&\times \left( \frac{b}{100~{\rm \mu m}} \right)\left( \frac{R_{\rm p}}{30~{\rm AU}} \right)^{3/2} \left( \frac{a}{500~{\rm AU}} \right)^{3/2}, \notag
\end{align}
where Eq. (\ref{eq:sigmagas}) is used for $\Sigma_{\rm gas}$ assuming $T_{\rm g} = 10^{4}$ K.  Because $t_{\rm drag}$ greatly exceeds the disk lifetime ($\max(t_{\rm visc}, t_{\rm fall})$; Eq.~\ref{eq:tvisc}), we may conclude that gas drag is negligible for debris evolution.  A possible exception occurs if the gas viscosity is quite low, with $\alpha \lesssim 10^{-3}$.  However, even if we assume that $t_{\rm drag}$ can be reduced to the the system age, the large majority of solid debris will not interact with the gas disk unless $f_{\rm rot} \sim 1$ (viscous spreading only brings a small percentage of the gas mass outward).  {\it The solid debris will most likely retain the highly eccentric trajectories of their cometary progenitors}.

\subsection{Debris Clouds: Observational Implications}

The above discussion establishes that, after an initial phase of dissipation and circularization, any gaseous accretion disk will drain rapidly before it has time to entrain or circularize significant amounts of solid debris.  The solid debris from cometary disruption will therefore remain on highly eccentric orbits, and will re-radiate absorbed WD luminosity at cooler temperatures.  This eccentric debris cloud will manifest observationally as an IR excess in the WD spectrum.  Such excesses are detected as point sources around $\sim 10\%$ of hot, young WDs \citep{Su+07, Chu+11}.  It is also possible that such emission could be resolvable by future high-resolution IR imaging (e.g. {\it JWST}).

To calculate the detailed properties of IR emission from the cloud, we project its luminosity density to produce a 2D surface brightness profile, $I(s)$.  We make the following assumptions in our calculation:
\begin{itemize}
\item The cloud is spherically symmetric.  This assumption is not entirely accurate, as the overlap between pre-kick and post-kick loss cones will preferentially place comets with orbital angular momenta parallel to the kick direction onto the lowest angular momentum trajectories.  Deviations from spherical symmetry grow with $v_{\rm k}$.  Analysis of our MC samples indicates that for kicks in the sweet spot of $v_{\rm k}\lesssim 1~{\rm km~s}^{-1}$, the sphericity assumption is correct to leading order, but is incorrect for significantly greater $v_{\rm k}$.
\item The cloud is optically thin.  This assumption is verified post-hoc for all models presented in this section.
\item Dust particles travel on closed Keplerian ellipses, such that emission at radius $r$ is calculated as the time-weighted average of every dust particle with a pericenter $r_{\rm p}<r$ and apocenter $r_{\rm a}>r$.
\item The dust is in thermal equilibrium with absorbed WD radiation at all times, so that $T_{\rm d}^4=L_{\rm WD}/(16\pi \sigma r^2)$, where $\sigma$ is the Stefan-Boltzmann constant and both emission and absorption cross-sections are geometric.  The corresponding Wien peak wavelength is generally smaller than the characteristic dust particle size set by radiation pressure blowout and PR drag ($2 \pi b \gtrsim \lambda$), justifying this assumption for $\lambda \lesssim 10^2~{\rm \mu m}$.
\item All dust particles have a semimajor axis distribution ${\rm d}N/{\rm d}a$ given by Eq. (\ref{aDist}).
\item All dust particles are spheres with a power-law distribution of radii $n(b) \propto b^{-L}$ for $L = 3.5$ as in Eq.~(\ref{dustPL}).
\end{itemize}
Under these assumptions, the bolometric luminosity density at the 3D radius $r$ is given by 
\begin{align}
j(r){\rm d}r=& \int^{\min(R_{\rm ice}, r)}_0{\rm d}r_{\rm p} \int_{\max(a_{\rm min},\frac{r_{\rm p}+r}{2})}^{\infty}{\rm d}a \frac{{\rm d}N}{{\rm d}r_{\rm p}{\rm d}{a}}\label{lumDensity} \\ 
&\times (1-f_{\rm gas}) \frac{M_{\rm OCA}\Upsilon(r_{\rm p})}{4 \pi r^2} \frac{L_{\rm WD}}{16 \pi r^2} \frac{{\rm d}t}{{\rm d}r} \frac{{\rm d}r}{\pi t_{\rm orb}} \notag
\end{align}
where $R_{\rm ice}$ is the maximum pericenter within which comets will sublimate (Eq.~\ref{rIce}), $M_{\rm OCA}$ is the total cometary mass in the OCA, and $t_{\rm orb}$ is the Keplerian orbital period for a given $a$ and $r_{\rm p}$.  The area-to-mass ratio $\Upsilon (r_{\rm p})$ is calculated self-consistently for each orbital pericenter $r_{\rm p}$, using Eq.~(\ref{eq:upsilon}) and the lower limit provided by non-gravitational radiation forces.  The distributions of $r_{\rm p}$ and $a$ are given by
\begin{equation}
\frac{{\rm d}N}{{\rm d}r_{\rm p}{\rm d}{a}} = \frac{{\rm d}N}{{\rm d}a} \frac{{\rm d}N}{{\rm d}r_{\rm p}},
\end{equation}
i.e.~the product of Eq. (\ref{aDist}) and the results of Figure \ref{massFunction}.  Furthermore, we use the Keplerian relation for elliptical orbits
\begin{equation}
\frac{{\rm d}r}{{\rm d}t} = \frac{na}{r}\sqrt{a^2e^2-(r-a)^2},
\end{equation}
where $n$ is the mean motion and orbital eccentricity $e=1-r_{\rm p}/a$.  The factor $\tilde{T}(r)=\pi^{-1}t_{\rm orb}^{-1}{\rm d}t/{\rm d}r$ appearing in Eq. (\ref{lumDensity}) is a small number that measures the fraction of time ${\rm d}t$ an elliptical orbit spends in the radial interval ${\rm d}r$, normalized such that $\int^{r_{\rm a}}_{r_{\rm p}} \tilde{T}(r){\rm d}r=1$, where $r_{\rm a}$ is orbital apocenter.

Projecting $j(r)$ along a line of sight to the 2D radial coordinate $s$ now gives the surface density profile 
\begin{equation}
I(s) = 2 \int^\infty_{\max(r_{\rm in}, s)} \frac{r j(r) {\rm d}r}{\sqrt{r^2-s^2}}
\end{equation}

\begin{figure}
\includegraphics[width=85mm]{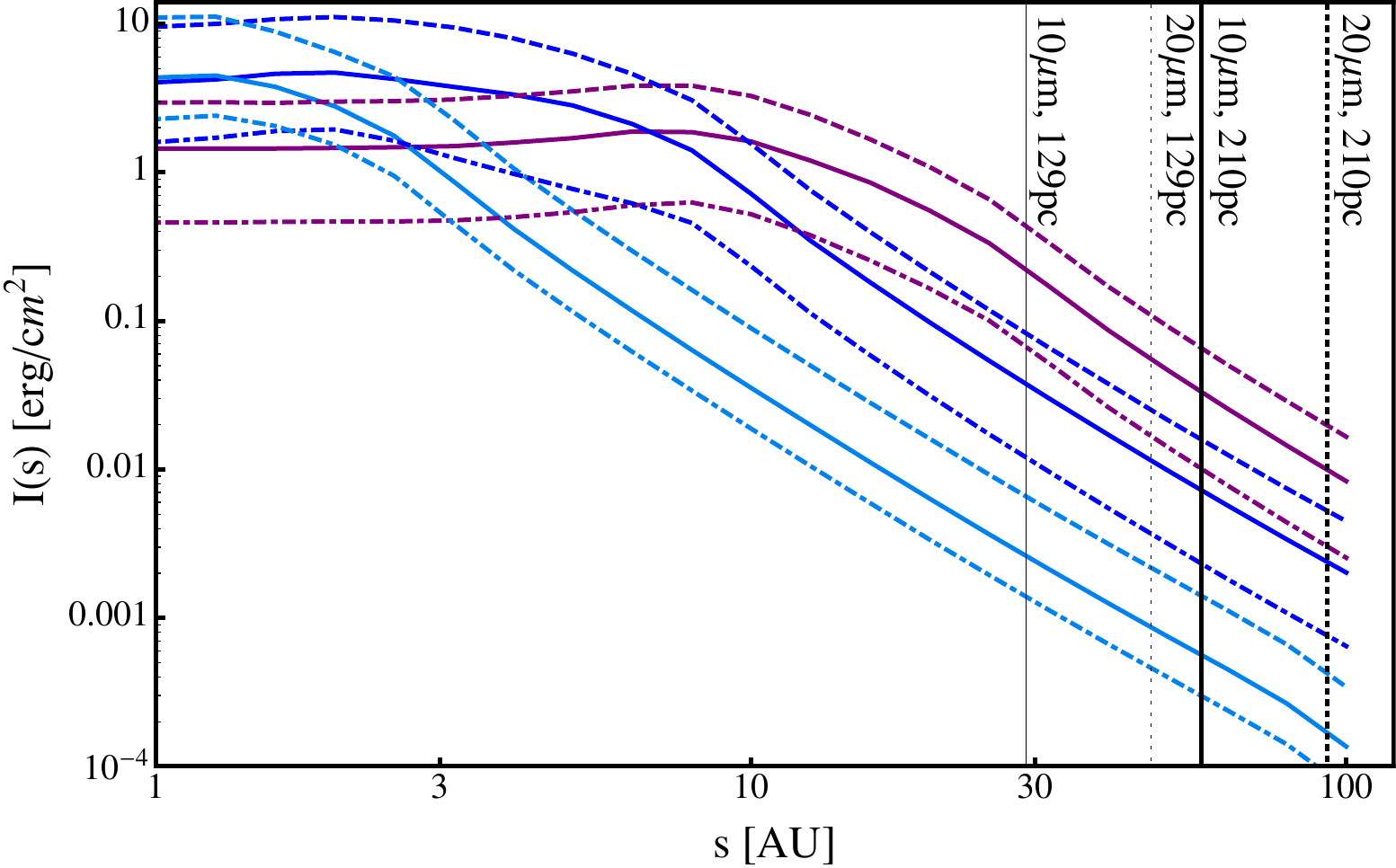}
\caption{Surface brightness $I(s)$ versus projected radial distance $s$, calculated for WD luminosities $L_{\rm WD}=100L_{\odot}$ (purple), $10L_{\odot}$ (blue), and $L_{\odot}$ (cyan).  Solid lines represent $v_{\rm k}=1~{\rm km~s}^{-1}$, dashed lines represent $v_{\rm k}=0.5~{\rm km~s}^{-1}$, and dot-dashed lines represent $v_{\rm k}=2~{\rm km~s}^{-1}$.  In all sets of curves we take initial minimum particle size $b_{\rm min}=0.1~{\rm \mu m}$, maximum particle size $b_{\rm max}=100~{\rm \mu m}$, and $M_{\rm OCA}=10M_{\oplus}$, although $b_{\rm min}$ is modified in a pericenter-dependent way to account for radiation forces.  The vertical black lines show diffraction-limited {\it JWST} angular resolution for the two nearest detected debris disks around young WDs, at both 10 and 20 $\mu$m.
}
\label{fig:cloudEmission}
\end{figure}

Figure \ref{fig:cloudEmission} shows the bolometric surface brightness profile calculated for WD luminosities of $1L_{\odot}, 10L_{\odot}$, and $100 L_{\odot}$, and kick velocities of $0.5~{\rm km~s}^{-1}$, $1~{\rm km~s}^{-1}$, and $2~{\rm km~s}^{-1}$.  In all of these profiles we take as fiducial parameters: $b_{\rm min} = 0.1 ~\mu m$; $b_{\rm max} = 100~\mu m$; OCA mass $M_{\rm OCA} = 10M_{\oplus}$; and time since the WD kick (important for calibrating dust depletion due to PR drag) of $10^5~{\rm yr}$.   In general, we find projected surface density profiles that peak between 1 and 10 AU and are relatively constant interior to that radius.  This constant-brightness core is created by a combination of projection effects and the truncation of the dust distribution at small radii by PR drag: the outer edge of the core corresponds to the radius where PR drag has eliminated almost all of the dust from our grain size distribution.

It is likewise straightforward to produce synthetic spectral energy distributions (SEDs) from our simulated clouds; the observed spectral flux density at Earth is given by $F_{\nu}=L_{\nu}/(4\pi d^2)$, with 
\begin{align}
L_{\nu} = 2\pi (1-f_{\rm gas}) M_{\rm OCA} & \int^{R_{\rm ice}}_{0}{\rm d}r_{\rm p} \int^\infty_{a_{\rm min}}{\rm d}a \int^{r_{\rm a}}_{r_{\rm p}}\Upsilon(r_{\rm p}) \notag \\
&\times B_{\nu}(r)\frac{{\rm d}N}{{\rm d}r_{\rm p}{\rm d}a}\frac{{\rm d}t/{\rm d}r}{\pi t_{\rm orb}} {\rm d}r, \label{Fnu}
\end{align}
where $d$ is the WD distance and $B_\nu (r)$ is the Planck function.

\begin{figure*}
\includegraphics[width=140mm]{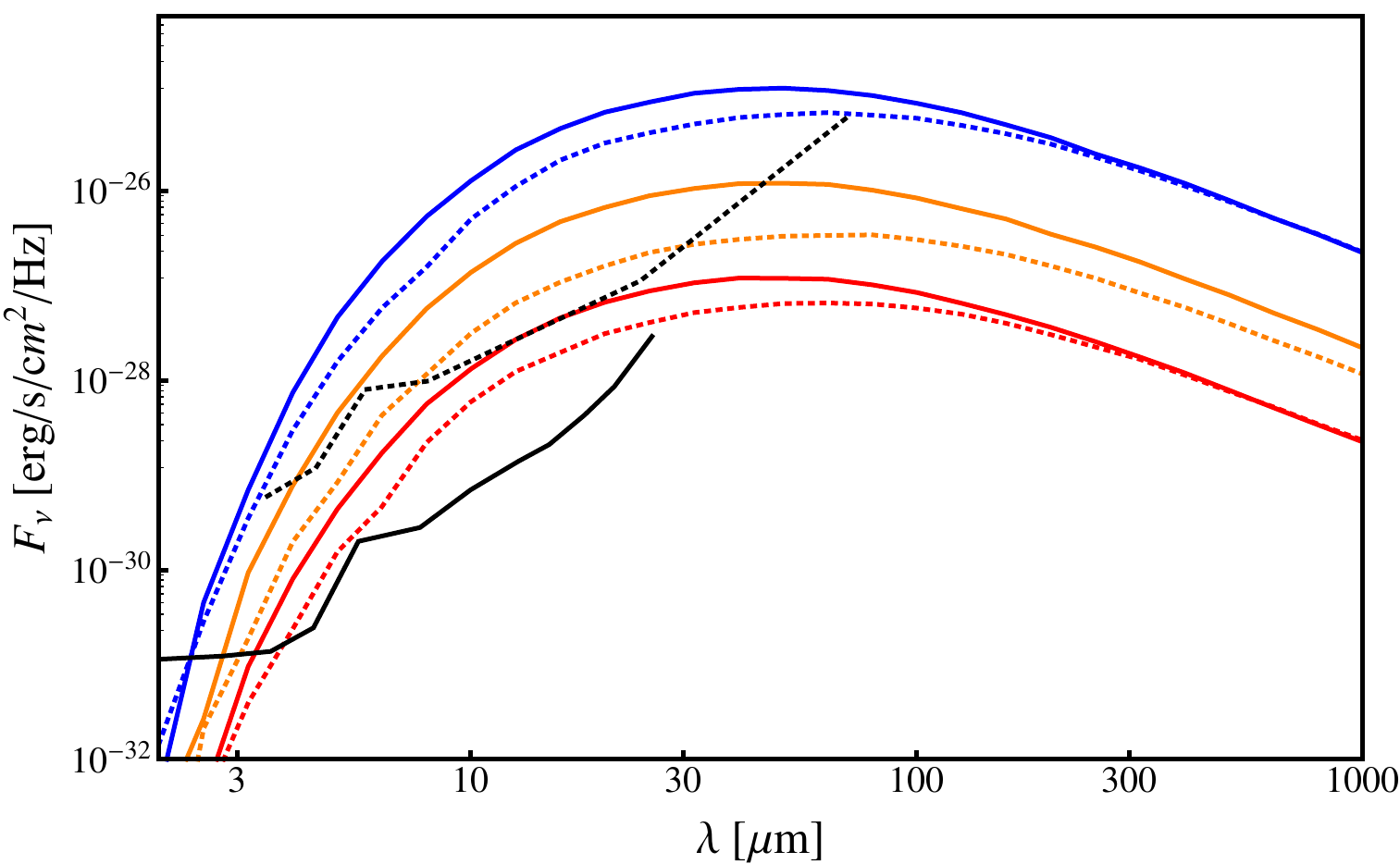}
\caption{Spectral flux density $F_{\nu}$ versus wavelength $\lambda$, calculated for two values of the WD luminosity $L_{\rm WD}=100L_{\odot}$ ({\it solid colored}) and $10L_{\odot}$ ({\it dotted colored}).  Blue lines show an `optimistic' scenario corresponding to an initial OCA mass $M_{\rm OCA}=50 M_{\oplus}$.  The orange and red curves represent `standard' and `pessimistic' scenarios for which $M_{\rm OCA} = 5 M_{\oplus}$ and $M_{\rm OCA} = 0.5 M_{\oplus}$, respectively.  All three scenarios assume an initial ice line (as determines the outer edge of the sublimated mass distribution ${\rm d}N/{\rm d}R_{\rm p}$) of $R_{\rm ice} = 30~{\rm AU}$, a kick velocity $v_{\rm k}=1~{\rm km~s}^{-1}$, and a distance $d=200~{\rm pc}$ corresponding to that of the Helix nebula.  The limiting sensitivities of {\it Spitzer} ({\it black dashed}) and {\it JWST} ({\it black solid}) are shown for comparison.}
\label{fig:cloudSED}
\end{figure*}

Figures \ref{fig:cloudSED} and \ref{fig:cloudSED2} show sample SEDs at distances of $200~{\rm pc}$, calculated for different assumptions regarding the mass $M_{\rm OCA}$, the WD luminosity, and the natal kick velocity $v_{\rm k}$.  Sensitivity curves for {\it Spitzer} and {\it JWST} are shown for comparison.  Since $F_{\nu} \propto M_{\rm OCA}\Upsilon$, detectability is sensitive to both OCA mass and to the distribution of sublimated dust grain sizes $n(b)$.  In Fig. \ref{fig:cloudSED} we can see that decreasing the WD luminosity moderately dims and reddens the resulting SED.  This reduction is less than might be expected, because the decreased temperature of the dust is partially compensated for by the increased value of $b_{\rm min}$.  Fig. \ref{fig:cloudSED2} shows the comparably modest dependence of the SEDs on $v_{\rm k}$.  Reasonable choices for $M_{\rm OCA}$ and $v_{\rm k}$ produce SEDs that are observable by {\it Spitzer} at a distance $200~{\rm pc}$ similar to that of the Helix, while {\it JWST} will be sensitive to clouds produced for a much wider range of parameters to significantly larger distances.  

\begin{figure*}
\includegraphics[width=140mm]{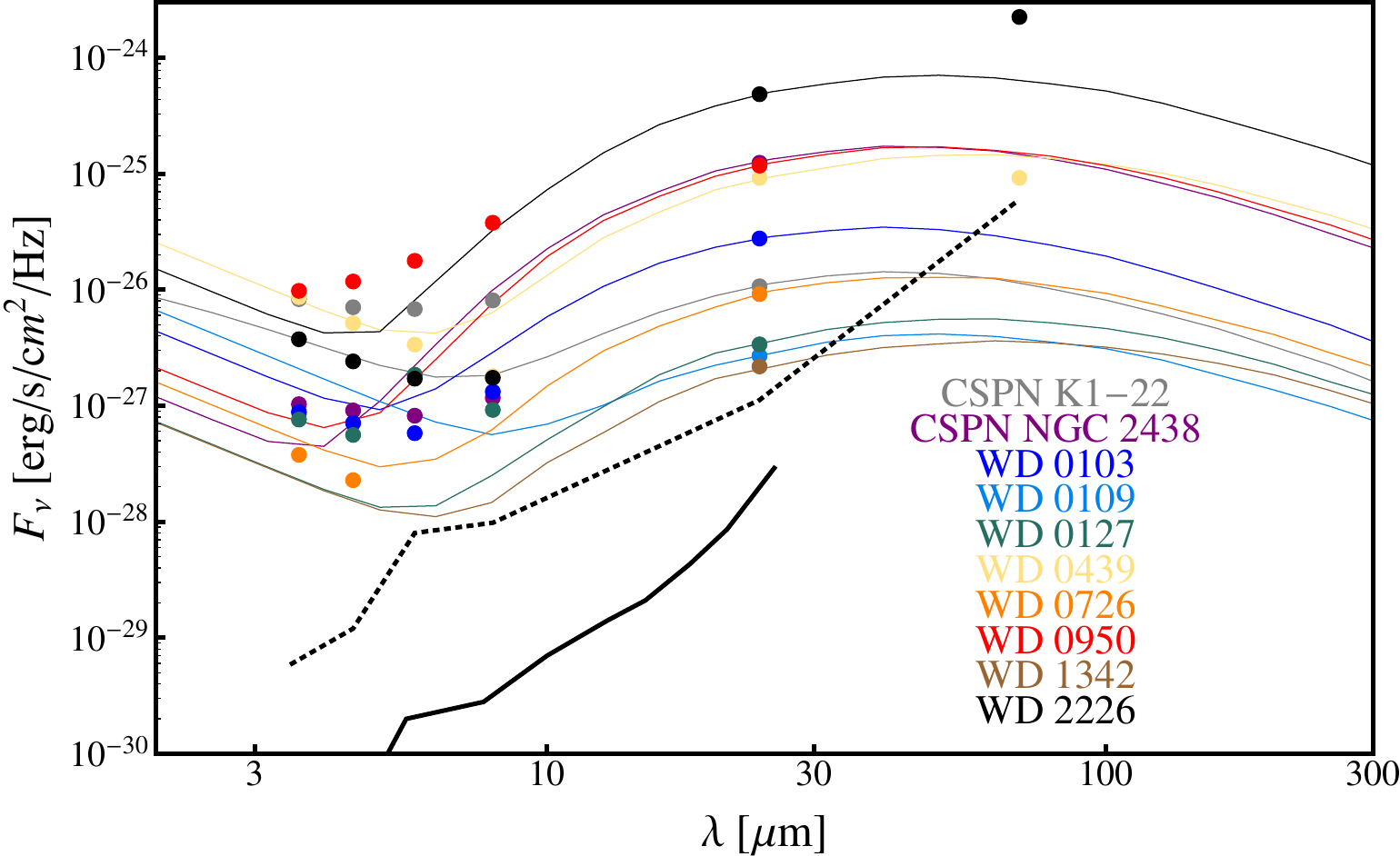}
\caption{Spectral flux density $F_{\nu}$ versus wavelength $\lambda$, calculated for all ten young WD systems observed to contain an IR excess by \citet{Chu+11}.  The spectral flux density curves include both dust emission from Eq. \eqref{Fnu} and blackbody emission from the central WD (and in the case of CSPN K1-22, blackbody emission from an unresolved companion star).  The primary free parameter is the total mass of the Oort Cloud analogue, $M_{\rm OCA}$, which is adjusted to fit the 24 $\mu$m data point.  The 3-8 $\mu$m data points are not fit for because of their sensitivity to time-dependent effects not accounted for in our model.  For five WD systems, this fitting gives Solar-like values for the total OCA mass, with $1.5 M_{\oplus} \lesssim M_{\rm OCA} \lesssim 20 M_{\oplus}$.  The other five systems require much larger OCA masses, with $M_{\rm OCA}/M_{\oplus}=\{145, 1225, 215, 365, 175\}$ for CSPN K1-22, CSPN NGC2438, WD 0103, WD 0950, and WD 2226, respectively.  As in prior plots, the {\it Spitzer} and {\it JWST} sensitivity curves are shown as dashed and solid black lines.}
\label{fig:cloudSEDOverlay}
\end{figure*}

\begin{figure}
\includegraphics[width=85mm]{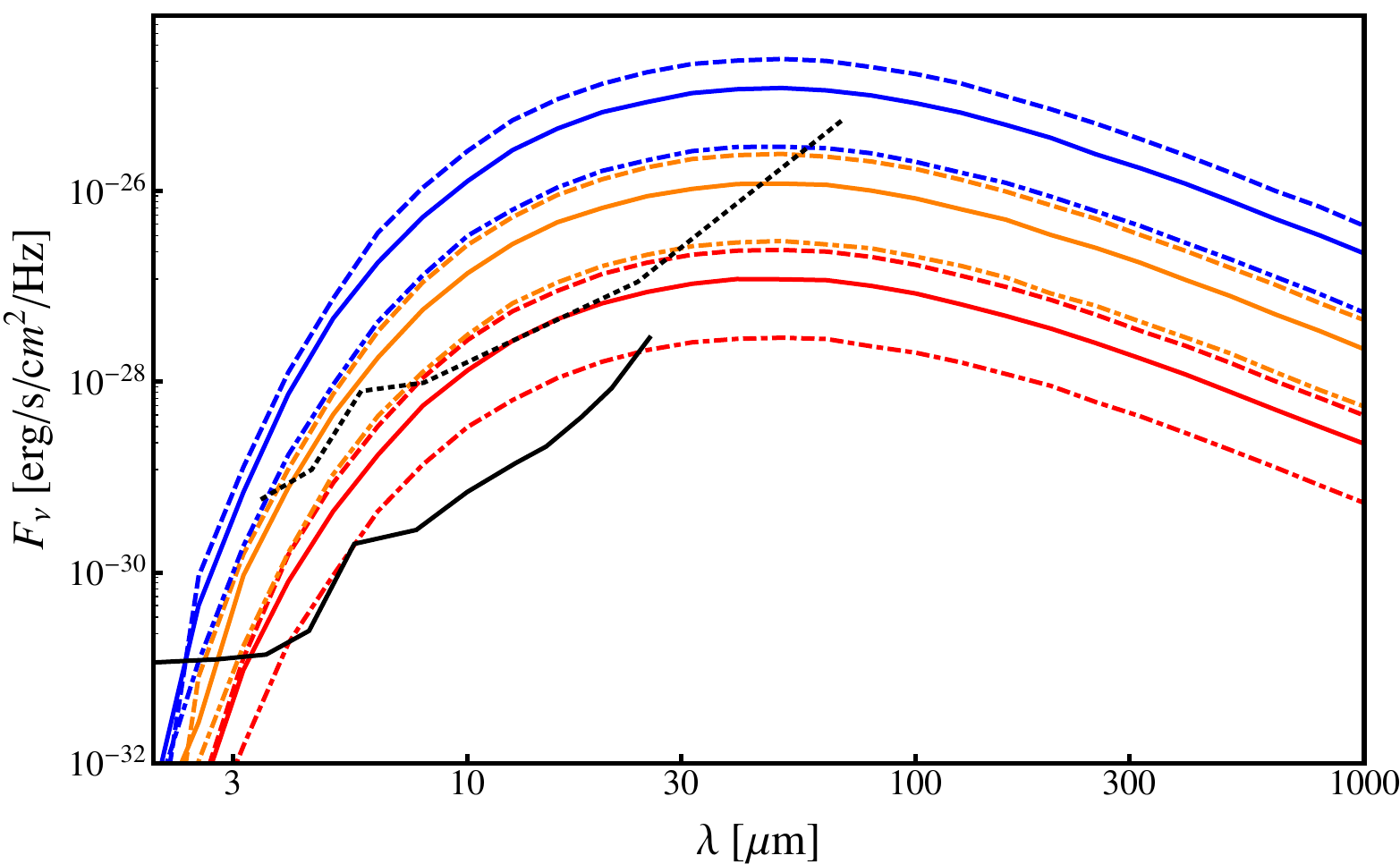}
\caption{Same as Figure \ref{fig:cloudSED}, but calculated for two different values for the strength of the WD natal kick $v_{\rm k}$ = 0.5 km s$^{-1}$ ({\it dashed colored lines}) and 2 km s$^{-1}$ ({\it dot-dashed colored lines}).  The colors, solid colored curves (fiducial model), and black sensitivity curves have the same meaning as in Figure \ref{fig:cloudSED}.}
\label{fig:cloudSED2}
\end{figure}

Fig. \ref{fig:cloudSEDOverlay} shows {\it Spitzer} data points for the ten young WDs\footnote{Two of these WDs are central stars of planetary nebulae (CSPN).} observed to contain significant IR excesses, in combination with predicted spectral flux densities $F_{\nu}$.  The plotted $F_{\nu}$ contains both dust emission from our model and the blackbody emission from the central WD; in both cases, $L_{\rm WD}$, $R_{\rm WD}$, and $d$ are taken from the observations described in \citet{Chu+11}.  The primary free parameter in our dust model is $M_{\rm OCA}$, which we fit to the 24 $\mu$m observations.  Kicks of $0.5~{\rm km~s}^{-1}$ are assumed.  In general, we are able to reproduce the observed long-wavelength IR excesses with reasonable values of Oort Cloud Analogue mass.  Specifically, we find that $24~{\rm \mu m}$ emission from CSPN K 1-22, CSPN NGC 2438, WD 0103+732, WD 0109+111, WD 0127+581, WD 0439+466, WD 0726+133, WD 0950+139, WD 1342+443, and WD 2226-210 can be reproduced with OCA masses (in $M_{\oplus}$) of roughly 145, 1225, 215, 2.0, 20, 26, 21, 365, 19, and 175, respectively.  Half of the observed systems are fitted with Solar-type masses of comets ($1.5 M_{\oplus} \lesssim M_{\rm OCA} \lesssim 20 M_{\oplus}$).  The other half require values of $M_{\rm OCA}$ an order of magnitude greater (almost two orders of magnitude for the exceptional case of CSPN NGC 2438), although three of these anomalously large $M_{\rm OCA}$ values may be artifacts of the extreme luminosities of the central WDs (CSPN K1-22, CSPN NGC 2438, and WD 0103), which make our model extremely sensitive to the upper cutoff in the assumed dust grain size distribution.

Several of the sources in the {\it Spitzer} sample have secondary $3-10~{\rm \mu m}$ IR peaks, which we do not fit for because of their sensitivity to time-dependent effects not accounted for in our exploratory model.  Specifically, at high frequencies dust emission will be:
\begin{enumerate}
\item Decreased by the higher temperatures of the WD at earlier times; comets sublimated at these times will see a much larger fraction of their debris eliminated by stronger PR drag forces.
\item Increased by the steady-state PR flow to small radii.
\end{enumerate}
Which of these effects wins out is not obvious, and would require development of a time-dependent PR drag model to estimate.  The four systems observed in \citet{Chu+11} which possess a $\sim 5 \mu$m emission bump may be indicative of high-temperature dust; we speculate that this may correspond to a steady-state PR flow at scales of $\sim 1 {\rm AU}$.  

\section{Discussion}

The {\it Spitzer} observations of ten young WDs possessing IR excesses were the primary motivation for the theoretical model produced in \S 4.  As discussed in \S 4.4, the combination of gentle WD kicks and Oort Cloud Analogues is capable of reproducing the $24~{\rm \mu m}$ {\it Spitzer} data points seen in all ten systems.  Half of these systems are well fit with $M_{\rm OCA}$ values in the Solar range ($2 \lesssim M_{\rm OCA}/M_{\oplus} \lesssim 30$), but the other half require OCA masses roughly an order of magnitude larger.  This may reflect selection biases, as the largest OCAs would produce the most luminous debris disks around young WDs.  However, our model is generally less capable of reproducing the hotter, second IR excesses seen in four of the {\it Spitzer} sample's WDs \citep{Chu+11}.  It is possible that this could be remedied by incorporating the time-dependent PR flow of dust into our model, or it could indicate that OCAs are unable to explain IR excesses around some fraction of young WDs.  We therefore summarize two alternative proposals in this section.

A commonly discussed explanation for debris disks around young WDs is a collisional cascade among KBO analogues \citep{Su+07}.  Such collisions may be triggered by the destabilization of planetesimal orbits due to interactions with surviving planets, whose orbits would themselves have been altered by mass loss from the central star \citep{Dong+10}.  Although {\it JWST} may help discriminate between the KBO scenario and our model as described above, other arguments disfavor the KBO hypothesis.  When the host star crosses the horizontal branch its luminosity reaches $\gtrsim 10^{3}L_{\odot}$ for a timescale of $\sim 10^7~{\rm yr}$.  During this epoch the water sublimation radius (Eq.~\ref{rIce}) moves outward to hundreds of AU, far beyond the Solar Kuiper Belt.  Although the comets in an OCA will generally orbit far enough away to survive, most volatile-dominated KBO analogues will likely be destroyed \citep{Stern+90} well before the AGB phase.\footnote{However, see \citet{JurXu10} for a discussion of how internal volatiles in minor planets can survive a host star's AGB phase.}  Large KBO analogues could survive sublimation in two ways: either by possession of a chemically differentiated, rocky interior, or by simply being large enough that their surface escape velocity $v_{\rm esc} \gg v_{\rm th}$.  Since a $L=6\times 10^3 L_{\odot}$ giant star will produce an equilibrium temperature $T_{\rm d} \approx 330~{\rm K}$ at $r=50~{\rm AU}$, the thermal velocity of water will be $v_{\rm th} \approx 0.7~{\rm km~s}^{-1}$ and sublimated ice will escape from even self-gravitating, Triton-sized objects.  Therefore, chemical differentiation and production of rocky cores offers the more promising mechanism for survival of very large KBO analogus.

Although the sublimation of KBO analogues will leave behind a debris ring of gas and dust (as in the case of cometary debris described in $\S\ref{sec:debris}$), the stellar luminosity will be much higher when this occurs.  If we assume a typical horizontal branch luminosity $\approx 6\times 10^{3}L_{\odot}$, then Eq. (\ref{radBlowout}) predicts that radiation pressure will remove all grains smaller than $2300 {\rm \mu m}$, while PR drag (Eq.~\ref{eq:tPR}) sets an even more stringent minimum size of $\approx 10^4~{\rm \mu m}$ \citep[see also \S 4 of][]{BonWya10}.  These lower limits for grain size are orders of magnitude above the $\approx 100\mu$m maximum grain size inferred based on the Deep Impact results for Tempel 1 ($\S\ref{sec:debris}$).  The KBO scenario may have serious difficulty producing enough dust that survives until the birth of the WD, unless surviving rocky cores have both a high enough spatial density to begin a collisional cascade, and sufficiently large total mass to reproduce observed IR luminosities.

A second alternative is the possibility that unresolved binary companions have created dust-rich disks during the WD's prior post-AGB phase \citep{Chu+11, Biliko+12}.  Analogous disks around post-AGB stars are generally associated with binarity \citep{vanWin+09}, but, as was pointed out by \citet{Chu+11}, circumstellar dust disks in post-AGB systems are typically too compact to produce an IR excess peaked at $24~{\rm \mu m}$ \citep{TaaRic10}.  Although they possess roughly correct temperatures for matching the IRAC-band ($3-8~ \mu$m) excess seen in $40\%$ of the \citet{Chu+11} sample, circumstellar disks generally have fractional luminosities $L/L_{\rm WD} \sim 0.3$ \citep{deRuyt+06}, orders of magnitude higher than the observed IRAC-band excesses, or for that matter the $24~\mu$m excesses.  It remains unclear if the hot, very luminous disks observed around post-AGB stars are able to survive in attenuated form into and past the PN stage.  During completion of this paper, new observational work appeared which indicates that 8 out of 13 CSPN stars with IR excesses also possess a binary companion \citep{Clayto+14}, strengthening the post-AGB disk scenario.

In the following subsections we discuss the observational implications of this paper's model for IR excesses around young WDs.  In addition to future IR observations of these dusty sources, we highlight more indirect implications as well, such as the atmospheric hydrogen content of old WDs, natal kick velocity distributions, and exoplanet survival during post-main sequence evolution.

\subsection{Future Observations of Disks Around Young WDs}
\label{sec:dustydisk}

Future observations by {\it JWST} of debris disks around young WDs may distinguish between our cometary hypothesis and competing theories involving objects from within the plane of the ecliptic such as KBO analogues (\citealt{Su+07}; \citealt{Dong+10}).  One `smoking gun' discriminant would be to spatially resolve the central IR cavity to distinguish a quasi-spherical cloud from one with a disk-like concentration (see discussion at end of $\S\ref{DD:ev}$).  Although a cavity of size $\sim 10{\rm AU}$ is predicted in both scenarios (due to a combination of radiation pressure blowout and PR drag), the cometary hypothesis predicts a relatively constant surface brightness profile due to projection effects (Fig.~\ref{fig:cloudEmission}) rather than the complete absence of emission predicted for a disk-like configuration observed at most inclinations.  

Though in principle a promising diagnostic, direct imaging of the central cavity will be challenging.  For the nearest young WD debris disk (distance $129~{\rm pc}$; \citealt{Chu+11}) $10~{\rm AU}$ corresponds to an angular size $0.16''$.  This is similar to the anticipated intrinsic resolution of the {\it Mid-Infrared Instrument (MIRI)} on {\it JSWT} ($0.11 '' / {\rm pixel}$), but the diffraction-limited resolution at $10 ~{\rm \mu m}$ ($20~{\rm \mu m}$)  is substantially greater $\approx0.45''$ ($0.89''$).  It is therefore unlikely that the central cavities can be resolved, although obviously the discovery of a significantly closer (and likely fainter) young WD debris disk would improve the prospects for a resolved source substantially.

A more feasible alternative discriminant might be large-scale imaging of these dusty IR sources.  Dust clouds produced by an OCA will generally appear circular in projection (unless $f_{\rm rot} \sim 1$, which is disfavored for the Solar Oort cloud).  However, if it is collisions between coplanar KBO analogues (or binary interactions) that produce the observed dust, then depending on the inclination angle of the debris disk, a range of ellipticities may result in projection.  In Fig. \ref{fig:cloudEmission}, we show the diffraction-limited angular resolution capabilities of {\it JWST} for the two nearest debris systems around young WDs at $10~{\rm \mu m}$ and $20~{\rm \mu m}$; if these systems can be imaged out to $\sim 50-100~{\rm AU}$, then their overall geometry can serve as a test of this paper's hypothesis.  

Unfortunately, at these distances, both $10~{\rm \mu m}$ and $20~{\rm \mu m}$ observations will probe the Wien tail of the dust's thermal emission.  For WD0439+466 (the 129~{\rm pc} system) the $10~{\rm \mu m}$ and $20~{\rm \mu m}$ values of $B_{\nu}$ are $4.9\times 10^{-3}$ and $1.4\times 10^{-1}$ of the blackbody peak, respectively.  For the Helix nebula (at $210~{\rm pc}$), these fluxes are reduced from the blackbody peak by factors of $3.6\times 10^{-3}$ and $1.6\times 10^{-3}$, respectively.  Observations at $20 \mu$m are therefore much more favorable.

The much greater spatial resolution of the Atacama Large Millimeter Array ({\it ALMA}) offers a possibly superior pathway for resolving the geometry of dusty IR sources around young WDs.  However, the SEDs we plot in Figs. \ref{fig:cloudSED}, \ref{fig:cloudSEDOverlay}, and \ref{fig:cloudSED2} are only accurate for wavelengths $\lambda \lesssim 100~{\rm \mu m}$.  A more accurate model for dust emission and absorption would be needed to accurately assess the observability of these debris clouds at {\it ALMA} wavelengths.

In addition to providing better data on known systems, {\it JWST} will be superb at detecting new WD debris disks.  A disk of luminosity $L_{\rm disk}$ at a distance $d$ will produce a flux
\begin{equation}
F_{\nu}=9600 ~{\rm \mu Jy}~ \frac{\Delta \lambda}{3 ~{\mu m}} \frac{L_{\rm disk}}{3\times10^{-3}L_{\odot}} \left(\frac{d}{100~{\rm pc}}\right)^{-2},
\end{equation}
where $\Delta \lambda$ is the bandpass, normalized to a value $3~\mu$m characteristic of {\it MIRI}.  Given the {\it MIRI} limiting sensitivity at $10~{\rm \mu m}$ ($20~{\rm \mu m}$) of $1 ~{\rm \mu Jy}$ ($9 ~{\rm \mu Jy}$), disks with SEDs peaking at these wavelengths will be detectable by {\it JWST} to a distance $\sim 10~{\rm kpc}$ ($\sim 1~{\rm kpc}$).  Even less luminous disks, as would be produced if the OCA is intrinsically low mass or if the WD kick is far from the optimal value $v_{\rm k}\sim 0.5~{\rm km~s}^{-1}$ (Fig.~\ref{massFunction}), are detectable to distances of hundreds of parsecs.

\subsection{Implications for WD Natal Kicks}
\label{sec:kick}
A variety of observations provide circumstantial, if not conclusive, evidence for WD birth kicks.  We briefly summarize these motivating observations, before discussing how observations of WD debris disks can substantiate the existence of natal kicks.

Young WDs in globular clusters may possess a more extended radial distribution than their older counterparts.  Such an effect was claimed in NGC 6397 \citep{Davis+08}, although recent {\it HST} observations have cast doubt on this conclusion \citep{Heyl+12}.  Weaker evidence exists for an extended young WD radial distribution in $\omega$ Cen \citep{Calami+08}.  If correct, these results are counterintuitive: WD progenitors are relatively massive stars expected to migrate to the center of their host cluster due to mass segregation, while WDs themselves are undermassive and hence should migrate back out of the center after forming.  Detailed dynamical modeling has found that such observations can be reproduced if a sizable fraction of WDs receive birth kicks $v_{\rm k}\sim \sigma$ \citep{Heyl07}, where $\sigma \sim few$ km s$^{-1}$ is the cluster velocity dispersion.  \citet{Fregea+09}, for example, find that $\sim 4~{\rm km~s}^{-1}$ kicks could reproduce observations of NGC 6397.  Importantly, the required kicks must also be impulsive with respect to the orbital timescale of the WD about the globular cluster, i.e. to occur on a timescale $\lesssim 10^5~{\rm yr}$.  Note that this condition is similar to that requiring the kick be impulsive with respect to the OCA ($\ref{sec:response}$), one of the key assumptions of our model.    


Open clusters, with velocity dispersions lower than those in globular clusters, offer in some ways a more promising arena to investigate weak WD natal kicks.  Open clusters are observed to possess a large depletion of WDs relative to their expected abundance from the IMF \citep{Weidem+92}, which birth kicks $v_{\rm k} \sim 1-5~{\rm km~s}^{-1}$ could resolve \citep{Fellha+03}.  This evidence is not conclusive, however, since alternative explanations for such WD deficits exist, most notably the hiding of WDs in binary systems.\footnote{Detailed Monte Carlo population modeling finds binarity could account for the observed WD deficit in the Pleiades, but not in the Hyades and Praesepe \citep{Willia04}.  However, more detailed recent observations of the Hyades appear to disfavor strong, $\gtrsim 1~{\rm km~s}^{-1}$ kicks \citep{SchRos11}.}

WD kicks may also help explain the large orbital eccentricies of WD-barium star binary systems, which are significantly higher than those expected from tidal circularization of the orbit during the AGB phase of the WD progenitor \citep{Izzard+10}. Asymmetric mass loss and velocity kicks during the AGB superwind phase provides a natural explanation for such high eccentricities \citep{Heyl07b}.  On the other hand, barium stars as a whole possess a large ($\approx 99\%$) binarity fraction, implying the kick velocity cannot be too high enough ($v_{\rm k} \lesssim 2-3~{\rm km~s}^{-1}$) or a large fraction of such systems would be dissociated (e.g.~\citealt{Izzard+10}).

Finally, WD natal kicks are a possible explanation for the short birth periods of WDs \citep{Spruit98}, which are otherwise difficult to explain given the expected magnetic coupling between between the WD-forming core and the envelope of the progenitor red giant. 

Given the critical role that WD kicks play in producing observational signatures of cometary sublimation (e.g. IR emission detected by {\it JWST}), a confirmation of our interpretation of such observations would likewise provide strong evidence for kick velocities in the range $0.1~{\rm km~s}^{-1}\lesssim v_{\rm k} \lesssim 2~{\rm km~s}^{-1}$.  As we now describe, observations of older, hydrogen-deficient WDs provide an independent (and possibly contradictory) probe of WD natal kicks.

\begin{figure*}
\includegraphics[width=155mm]{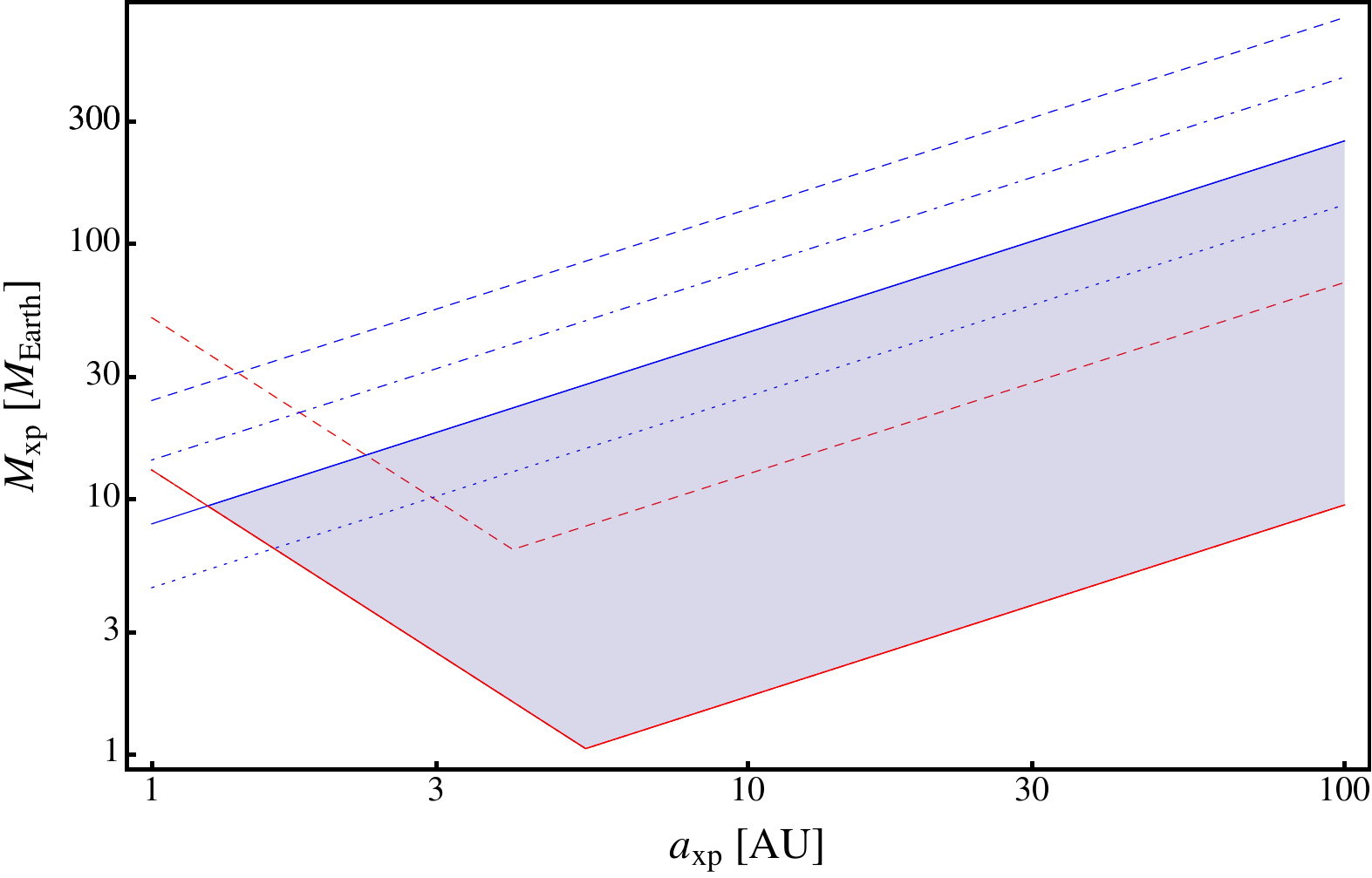}
\caption{Parameter space of exoplanet properties (mass $M_{\rm xp}$ and semi-major axis $a_{\rm xp}$) conducive to OCA formation (see Appendix A for details).  Solid lines represent the minimum ({\it red}) and maximum ({\it blue}) planet mass necessary to perturb a belt of planetesimals into a long-lived Oort cloud analogue, calculated for a solar type star of mass $M_*=M_{\odot}$ and age $t_*=10^{10}~{\rm yr}$ embedded in a region of stellar density $\rho_* = 0.15M_{\odot}~{\rm pc}^{-3}$ similar to that of the Solar neighborhood.  The region of parameter space capable of producing an OCA is shaded for this fiducial scenario.  The dashed lines show an otherwise identical model calculated for a a more massive 2.5$M_{\odot}$ star.  Dotted and dot-dashed lines show the maximum mass calculated for a $M_{\odot}$ star in regions of lower ($\rho_* = 0.015M_{\odot}~{\rm pc}^{-3}$) and higher ($\rho_* = 1.5M_{\odot}~{\rm pc}^{-3}$) ambient stellar density, respectively, than the fiducial case (the lower bound on exoplanetary mass is unchanged from the fiducial model shown with a solid line). 
}
\label{fig:OortScenarios}
\end{figure*}

\subsection{Hydrogen Abundances in Old WDs}
Roughly half of the DB WDs in the \citet{Berger+11} sample\footnote{DB stars make up 64/108 stars in this paper; DBA stars make up the remaining 44/108.  DB stars of both types are, however, a minority of all WDs \citep{Eisens+06}.} have so little atmospheric hydrogen that their observed spectra are incompatible with accretion of a single modestly sized comet, and are many orders of magnitude below our estimates for total accreted $M_{\rm H}$.  Even most of the DBA atmospheres in \citet{Berger+11} have hydrogen abundances typically one or more orders of magnitude below our fiducial $M_{\rm H} \sim 5\times 10^{23}~{\rm g}$ estimates.

These observed hydrogen deficiencies could, theoretically, be explained in the following ways:
\begin{enumerate}
\item Convective dilution of a small surface hydrogen layer once the WD cools below a certain temperature.
\item Extremely late-time surface burning of hydrogen.
\item Loss of surface hydrogen in a line-driven wind.
\item An unfavorably low OCA mass.
\item A significantly small, large, or non-impulsive WD natal kick.
\end{enumerate}
Although option (i) is widely believed to explain the observed DB deficiency among older, cooler WDs \citep{FonWes87}, it manifestly cannot explain the observed hydrogen absences in the younger, hotter DB population discovered more recently \citep{Eisens+06}.  Option (ii), the so-called ``born again'' scenario held to be responsible for Sakurai's Object \citep{Duerbe+00, Herwig+11}, invokes a Very Last Thermal Pulse (VLTP) of hydrogen burning after the star has left the AGB branch, briefly reinflating the star to a large size.  However, this VLTP is short-lived, and is generally taken to occur $\sim 10^5~{\rm yr}$ after the star leaves the AGB track \citep{Iben+83}.  Although a H-burning pulse at these times could wipe out the hydrogen accreted from many, perhaps most, of the comets in our MC samples, a fraction of these comets come from semimajor axes $a \gtrsim 10^4 ~{\rm AU}$ and therefore do not arrive at the WD until $\sim 10^6~{\rm yr}$ after the natal kick.  Given the enormous value of $M_{\rm H}$ relative to \citet{Berger+11} limits, option (ii) seems unlikely.  Option (iii) is also not promising; the high surface gravity of WDs poses a significant barrier to launching line-driven winds, and even if they can be driven at low mass loss rates, they will not collisionally couple to hydrogen and helium \citep{Unglau08}.

The remaining possibilities, options (iv) and (v), are of much astrophysical interest, as they touch on important open questions: the existence of OCAs and WD natal kicks.  The formation of the Solar Oort Cloud depended critically on the existence of Solar System gas giants; without a properly configured gas giant system to scatter planetesimals, it is possible that a subset of exoplanetary systems would lack OCAs entirely, or at least see serious reductions in their mass (see $\S\ref{sec:exo}$ for further discussion).  

Option (v) also has some plausibility: impulsive natal kicks $v_{\rm k} \gtrsim 4~{\rm km~s}^{-1}$ are found to significantly reduce $M_{\rm evap}$ (Fig. \ref{periBoth}).  However, we again note that an enormous reduction in $M_{\rm evap}$, much larger than is seen for a $4~{\rm km~s}^{-1}$ kick, is needed to explain some of the observed DB hydrogen fractions.  Furthermore, kicks much larger than this value are significantly constrained by observations of WDs in globular clusters (\S 5.2).  A very small natal kick, $v_{\rm k} \ll 0.1~{\rm km~s}^{-1}$, could also shut off the hydrogen delivery mechanism proposed in this paper; a non-impulsive natal kick of larger magnitude would have the same effect.  Although some combination of these possibilities appears to be the cleanest explanation of the low DB hydrogen fractions, we note that a prediction in this cases is the survival of a massive OCA.  Over much longer timescales, perturbations to this OCA by the galactic tide or passing stars could lead to the accretion of a handful of comets and the deposition of significant $M_{\rm H}$ \citep{WieTre99}; however, calculating typical rates of comet accretion from these mechanisms is beyond the scope of this paper.

An important point raised in $\S\ref{sec:kick}$ concerns the survival of OCAs in dense stellar environments.  Because WDs residing in globular or sufficiently dense open clusters will lose their OCAs over time, the floor for atmospheric hydrogen abundances will not be set by cometary accretion but instead by ISM accretion.  A key prediction of our model is therefore that the ratio of extremely H-deficient DBs to the overall DB population should be greater in globular (and perhaps dense open) clusters.  This argument is strengthened by the low metallicities of globular cluster stars, which likely impeded or prevented the formation of OCAs.

\subsection{Implications for Exoplanetary Systems}
\label{sec:exo}
The production of OCAs requires relatively massive planets orbiting at substantial distances from their host stars.  The schematic picture we present in Appendix A is largely based on \citet{Tremai93} and describes how pericentric interactions with an interior planet will perturb the orbital energies of a coplanar planetesimal belt.  Following a gradual process of energy diffusion, orbital evolution is taken over by the galactic tide, which leads to angular momentum diffusion at fixed energy, and isotropization of cometary inclinations.  One notable feature of this model is that it requires a relatively restricted range of planetary parameters (mass $M_{\rm xp}$ and semi-major axis $a_{\rm xp}$) in order to form an OCA.  A sizeable (at least super-Earth in mass) exoplanet must reside at a semimajor axis $\gtrsim 1~{\rm AU}$.  There is therefore a reasonable chance that any young WD with evidence of a surrounding OCA would possess an exoplanet that survived its host star's post-main sequence evolution, particularly if the progenitor star was not too massive \citep{VilLiv07}.

Figure \ref{fig:OortScenarios} shows the allowed parameter space of exoplanetary systems that will allow the production of an OCA around a solar mass star.  Generally, a Neptune- or Uranus-like exoplanet is required, although at small semimajor axis, super-Earths may suffice.  The allowed mass range vanishes for $a_{\rm xp} \lesssim 1~{\rm AU}$, as planetesimals then collide with the perturbing planet before the galactic tide can begin to alter their pericenter.  Stars with a greater zero-age main sequence mass require larger planets; a $2.5M_{\odot}$ star requires Uranus- to Saturn-sized exoplanets to form an OCA.

If our interpretation of the infrared excesses around young WDs is correct, then the $\sim 15\%$ detection rate of 24$\mu$m excess (\citealt{Chu+11}) implies an equally large fraction of exoplanetary systems possess a middle to large-weight gas giant with $a \gtrsim {\rm few}$ AU.  WD debris disks may thus provide a unique probe of such objects, which are are otherwise challenging to detect with radial velocity or transit surveys.  This number is broadly compatible with inferred abundances of large-separation planets from microlensing surveys \citep{Cassan+12}, which are more sensitive to this type of exoplanet.  However, given the existence of competing explanations (both dynamically active KBO analogues, and AGB wind fallback disks) for the IR excesses discussed in this paper, more followup work is needed before firm conclusions can be drawn about surviving exoplanets in such systems.

\section{Conclusions}
We have examined the response of extrasolar Oort cloud analogues to the kicks that may accompany WD birth.  Under the assumption that WDs receive a modest, $\sim 1~{\rm km~s}^{-1}$ natal kick over a sufficiently short timescale ($\lesssim 10^5~{\rm yr}$), then $\sim 10^{-3}$ of the mass of the associated OCA will be placed onto nearly radial orbits resulting in sublimation near pericenter.  This has several important observational implications, which we list here:
\begin{itemize}
\item The solid debris from the sublimation process will form a roughly spherical, optically thin cloud around the WD, with dust grains absorbing and re-radiating a fraction of the WD's luminosity.  The resulting IR emission (Figs.~\ref{fig:cloudEmission}, \ref{fig:cloudSED}, \ref{fig:cloudSED2}) can be fit into rough agreement with the $24~\mu {\rm m}$ IR excesses observed by {\it Spitzer} around $\sim 15\%$ of newborn WDs.  For roughly half the sample this can be accomplished with Oort Cloud Analogue masses comparable to the mass of the Solar Oort Cloud ($M_{\rm OCA}\sim 10~M_{\oplus}$), but for the other half of the sample values of $M_{\rm OCA} > 100~M_{\oplus}$ are needed.
\item Roughly half of young WD debris disks are seen to possess a second IR excess at much shorter ($\sim 3~{\rm \mu m}$) wavelengths, which is not well fit by our model.  The simple model presented in this paper is a time-independent one that excises all dust removed from the system by Poynting-Robertson drag; we plan to investigate this steady state, small-radius PR flow in future work to determine whether it could produce a second IR peak.  It is also possible that the systems which exhibit the short-wavelength IR peak acquire both dust excesses through an alternate mechanism, e.g. an unresolved stellar binary companion.
\item Our proposed mechanism for producing the observed IR excess differs from past hypotheses primarily in its geometric structure: spherical, rather than disk-like.  Our mechanism also employs a progenitor population residing far enough from the WD to resist sublimation during post-main sequence evolution; more tightly bound populations of planetesimals (e.g. Kuiper Belt analogues) are likely to sublimate before the planetary nebula phase.  
\item Young WDs with evidence for OCAs are also likely to possess sizeable planets orbiting at large enough semimajor axis to have survived post-main sequence evolution (Fig.~\ref{fig:OortScenarios}).  These WDs could be attractive targets for planet searches.
\item The very late heavy bombardment predicted in this paper is at odds with the hydrogen fraction in extremely H-deficient DB WDs.  Many DB WDs have less hydrogen in their atmospheres than would be delivered by the accretion of a single moderately sized comet, and therefore must have either {\it (i)} received a birth kick $v_{\rm k} \ll 1~{\rm km~s}^{-1}$ or $v_{\rm k} \gg 1~{\rm km~s}^{-1}$; or {\it (ii)} not have possessed an OCA.
\end{itemize}

Although this exploratory study has described the basic features of debris clouds produced around young WDs by natal kicks, several important theoretical problems remain for future work.  These include time-dependent models for dust inspiraling due to PR drag; a realistic model for frequency-dependent dust absorption and emission; and a dynamical study of comet accretion onto WDs in the absence of a natal kick, which will test whether the extremely H-deficient DB WDs can only be explained by option {\it (ii)}: a complete absence of OCAs.

For the nearest young WDs with IR excesses, it is possible that future imaging by {\it JWST} will provide a smoking gun test for this hypothesis, either by detecting the central cavity (due to PR drag) characteristic of disk scenarios, or by directly measuring the shape of the emitting area at larger sizes (true disks will appear non-circular due to inclination with respect to our line of sight).  Careful spectroscopy may also be able to distinguish between these hypotheses.  However, even with current {\it Spitzer} observations, the OCA mechanism proposed in this paper has one strong advantage over KBO scenarios: its cometary reservoirs of dust and gas survive the post-main sequence evolution of the parent star, which is not obviously the case for Kuiper Belt analogues.  Current and future observations of debris disks around young WDs may therefore offer a rare opportunity to probe distributions of comets in extrasolar planetary systems.

\section*{Acknowledgments}
We would like to thank John Debes, Jay Farihi, John Fregeau, Mike Jura, Chris Matzner, Norm Murray, Roman Rafikov, and Yanqin Wu for helpful discussions.  We would also like to thank the anonymous referee for many useful suggestions that have improved this paper.  A.L. was supported in part by NSF grant AST-1312034.  BDM gratefully acknowledges support from the NSF grant AST-1410950 and the Alfred P. Sloan Foundation.

\appendix

\section{OCA Formation}

A detailed analysis of OCA formation is beyond the scope of this paper, but we outline here the relevant physical mechanisms, and illustrate the dependence of OCAs on both their originally associated exoplanets, and their position in the galaxy.  The simple treatment given here is taken largely from \citet{Tremai93}.  We take as simplified initial conditions a recently formed exoplanet of mass $M_{\rm xp}$, radius $R_{\rm xp}$, and semimajor axis $a_{\rm xp}$ on a circular orbit, with a disk of planetesimals both interior and exterior to it.  Some of these planetesimals will have planet-crossing orbits, which will lead to orbital diffusion in energy (angular momentum is approximately conserved because all energy diffusion occurs near pericenter).  Using the energy-like variable $x=a^{-1}$, the diffusion coefficient $D_{\rm x}=\langle(\Delta x)^2\rangle^{1/2}$ is given by
\begin{equation}
D_{\rm x}=\frac{10M_{\rm xp}}{a_{\rm xp}M_*},
\end{equation}
where $M_*$ is the main-sequence mass of the central star.  The diffusion timescale for (order unity changes in) the semimajor axis of a planetesimal with orbital period $P$ and dimensionless energy $x$ is 
\begin{eqnarray}
&& t_{\rm diff} = P\frac{x^2}{D_{\rm x}^2} \nonumber \\
&& \approx 1.1\times10^9~{\rm yr} \left(\frac{M_*}{M_{\odot}} \right)^{3/2} \left( \frac{M_{\rm xp}}{M_{\oplus}} \right)^{-2} \left( \frac{x}{\rm AU^{-1}} \right)^{1/2} \left( \frac{a_{\rm xp}}{\rm AU} \right)^2. \nonumber \\
\end{eqnarray}
If we assume that most of the planet-crossing planetesimals have $x \approx a_{\rm xp}^{-1}$, then diffusive evolution within the main sequence lifetime of the central star, $t_*$, requires
\begin{align}
&\frac{M_{\rm xp}}{M_{\oplus}} \gtrsim \left( \frac{M_*}{M_{\odot}} \right)^{5/7} \left( \frac{t_*}{10^9~{\rm yr}} \right)^{-1/2} \left( \frac{a_{\rm xp}}{\rm AU} \right)^{3/4} \label{constraint1} \\
&\frac{M_{\rm xp}}{M_{\oplus}} \gtrsim 0.3\left( \frac{M_*}{M_{\odot}} \right)^{2.19}  \left( \frac{a_{\rm xp}}{\rm AU} \right)^{3/4}. \notag
\end{align}
The second of these inequalities has replaced $t_*$ with the main sequence stellar lifetime $T_*=10^{10}~{\rm yr}~(M_*/M_{\odot})(L_{\odot}/L_*)$, using the upper main sequence relation $L_* \propto M_*^{3.88}$.  After a time $t_{\rm diff}$, planetesimals will begin escaping the potential well of the central star into unbound orbits.  However, this escape can be halted by the effect of the galactic tide, which alters the angular momentum of the planetesimals' orbits.  Once the tidal timescale,
\begin{eqnarray}
&&t_{\rm tide} \sim  \nonumber \\
&&10^{15}~{\rm yr}\left( \frac{M_*}{M_{\odot}} \right)^{1/2} \left( \frac{\rho_*}{0.15M_{\odot}{\rm pc}^{-3}} \right)^{-1} \left( \frac{r_{\rm p}}{\rm AU} \right)^{1/2} \left( \frac{x}{\rm AU^{-1}} \right)^2,\nonumber \\
\end{eqnarray}
becomes comparable to $t_{\rm diff}$, the orbital pericenter of a planetesimal's orbit will shift and the perturbations from the exoplanet will cease to be relevant.  Subsequently the planetesimal (which we will hereafter refer to as a comet) will diffuse through angular momentum space at fixed energy under the influence of the galactic tide; the opposite of its earlier energy diffusion at fixed angular momentum.  In the above equation $r_{\rm p}$ is the orbital pericenter of the comet, and $\rho_*$ is the spatial density of surrounding stars.

By equating $t_{\rm tide} \sim t_{\rm diff}$, we can calculate the final semimajor axis at which a comet's orbital energy freezes out.  This occurs at 
\begin{eqnarray}
&&a_{\rm f} \approx  1\times 10^4{\rm AU} \times \nonumber \\
&&\left( \frac{M_*}{M_{\odot}} \right)^{-2/3} \left( \frac{\rho_*}{0.15M_{\odot}{\rm pc}^{-3}} \right)^{-2/3} \left( \frac{M_{\rm xp}}{M_{\oplus}} \right)^{4/3} \left( \frac{a_{\rm xp}}{\rm AU} \right)^{-1} \nonumber \\
\end{eqnarray}
provided that $D_{\rm x} \lesssim 1/a_{\rm f}$.  If this condition is not satisfied, a large majority of the comets will diffuse onto unbound orbits before their energy can freeze out at a bound value.  This condition allows us to put a second constraint on the perturbing exoplanet, namely
\begin{equation}
\frac{M_{\rm xp}}{M_{\oplus}} \lesssim \left( \frac{M_*}{M_{\odot}} \right)^{5/7} \left( \frac{\rho_*}{0.15~M_{\odot}~{\rm pc}^{-3}} \right)^{2/7} \left( \frac{a_{\rm xp}}{\rm AU} \right)^{6/7}. \label{constraint2}
\end{equation}
However, there are three further effects which can prevent or extinguish the existence of an OCA: galactic tidal stripping, direct collisions with the perturbing exoplanet, and encounters with passing stars.  We will briefly consider each of these so as to better delineate the exoplanetary parameter space conducive to OCA formation and survival.  

The central star's gravitational influence does not dominate to an infinite distance, but is instead truncated at a tidal radius
\begin{equation}
a_{\rm t} = 1.7\times 10^5~{\rm AU} \left(\frac{M_*}{M_{\odot}} \right)^{1/3} \left( \frac{\rho_*}{0.15~M_{\odot}~{\rm pc}^{-3}}\right)^{-1/3},
\end{equation}
beyond which comets will be stripped by the galactic tide.  To prevent the tidal stripping of most comets, we require $a_{\rm f} \lesssim a_{\rm t}$, or equivalently
\begin{equation}
\frac{M_{\rm xp}}{M_{\oplus}} \lesssim 8 \left( \frac{M_*}{M_{\odot}} \right)^{3/4}  \left( \frac{\rho_*}{0.15~M_{\odot}~{\rm pc}^{-3}}\right)^{1/4} \left( \frac{a_{\rm xp}}{\rm AU} \right)^{3/4}. \label{constraint3}
\end{equation}
We also must require the (rarely stringent) condition that $a_{\rm f} \gtrsim a_{\rm xp}$ in order for our freeze-out analysis to apply; this implies 
\begin{equation}
\frac{M_{\rm xp}}{M_{\oplus}} \gtrsim 10^{-3} \left( \frac{M_*}{M_{\odot}} \right)^{1/2}  \left( \frac{\rho_*}{0.15~M_{\odot}~{\rm pc}^{-3}}\right)^{1/2} \left( \frac{a_{\rm xp}}{\rm AU} \right)^{3/2}. \label{constraint4}
\end{equation}
In order to diffuse to the freeze-out energy, a planetesimal must avoid direct impacts on the perturbing exoplanet.  Because $N=(x_{\rm xp}/D_{\rm x})^2$ orbits are required in order to reach energy freeze-out, and the per-orbit impact probability (neglecting gravitational focusing and mean motion resonances) is $P_{\rm i} = (R_{\rm xp}/a_{\rm xp})^2/\Delta\theta$, where $\Delta \theta$ is the inclination thickness of the planetesimal disk, we can rewrite the requirement that $NP_{\rm i} \lesssim 1$ as 
\begin{equation}
\frac{M_{\rm xp}}{M_{\oplus}} \gtrsim 13 \left(\frac{M_*}{M_{\odot}} \right)^{3/2} \left( \frac{a_{\rm xp}}{\rm AU} \right)^{-3/2} \left( \frac{\rho_{\rm xp}}{3~{\rm g~cm}^{-3}} \right)^{-1/2} \left( \frac{\Delta \theta}{0.1} \right)^{-3/4}. \label{constraint5}
\end{equation}
This requirement seriously limits the ability of small exoplanets, or exoplanets on tightly bound orbits, from generating an OCA.  

Finally, we must require that encounters with passing stars do not dissipate a successfully formed OCA within the host star's main sequence lifetime.  The half-life of Oort cloud comets to stellar perturbations is 
\begin{equation}
t_{1/2} = 10^{10}~{\rm yr} \left( \frac{M_*}{M_{\odot}} \right)  \left( \frac{\rho_*}{0.15~M_{\odot}~{\rm pc}^{-3}}.\right)^{-1} \left( \frac{a}{10^4~{\rm AU}} \right)^{-1}
\end{equation}
Most comets will survive these perturbations so long as $t_{1/2} \lesssim t_*$, or, equivalently, 
\begin{align}
&\frac{M_{\rm xp}}{M_{\odot}} \lesssim 6 \left( \frac{M_*}{M_{\odot}} \right)^{5/4} \left( \frac{t_*}{10^9~{\rm yr}} \right)^{-3/4}  \left( \frac{\rho_*}{0.15~M_{\odot}~{\rm pc}^{-3}}\right)^{-1/4} \left( \frac{a_{\rm xp}}{\rm AU} \right)^{3/4} \label{constraint6} \\
&\frac{M_{\rm xp}}{M_{\odot}} \lesssim 1.1 \left( \frac{M_*}{M_{\odot}} \right)^{3.41}  \left( \frac{\rho_*}{0.15~M_{\odot}~{\rm pc}^{-3}}\right)^{-1/4} \left( \frac{a_{\rm xp}}{\rm AU} \right)^{3/4} 
\end{align}
In the latter inequality, we have again replaced $t_*$ with $T_*$.  

\begin{figure}
\includegraphics[width=85mm]{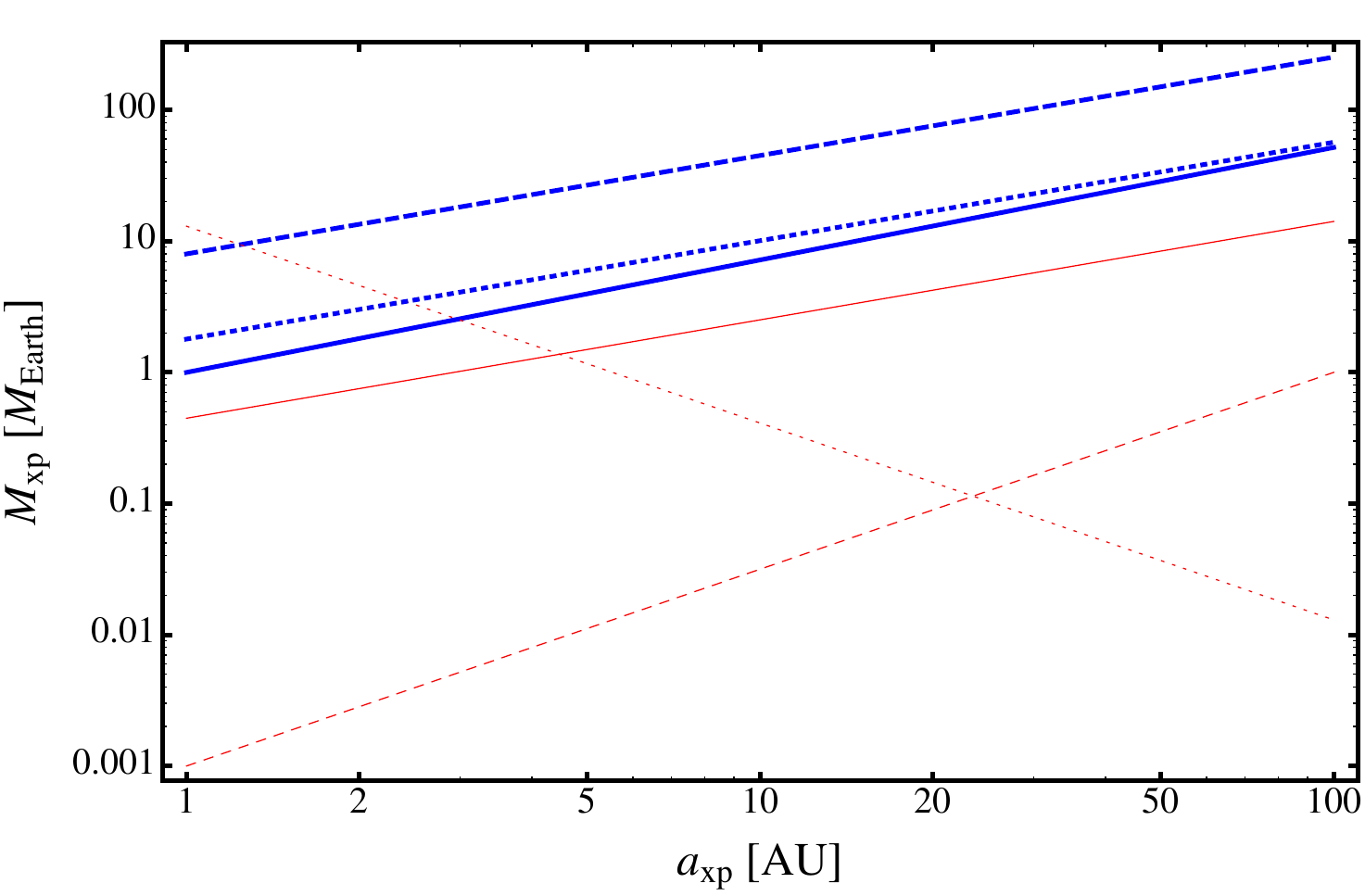}
\caption{Constraints on the formation of Oort cloud analogues by exoplanets of mass $M_{\rm xp}$ and orbital semimajor axis $a_{\rm xp}$.  In this fiducial case we take a solar-type star in the solar neighborhood of the galaxy, with $M_*=M_{\odot}$ and $\rho_* = 0.15M_{\odot}~{\rm pc}^{-3}$; we plot these constraints at a time $t_*=5\times 10^9~{\rm yr}$ after its formation.   We furthermore take the exoplanet's density $\rho_{\rm xp}=3~{\rm g}~{\rm cm}^{-3}$, and consider a belt of planetesimals of initial thickness $\Delta \theta=0.1$.  The thin red curves represent lower bounds on the required planetary mass, while the thick blue curves represent upper bounds.  The solid thin curve, solid thick curve, thick dashed curve, thin dashed curve, thin dotted curve, and thick dotted curve denote Eqs. \ref{constraint1}, \ref{constraint2}, \ref{constraint3}, \ref{constraint4}, \ref{constraint5}, and \ref{constraint6}, respectively.  The most constraining requirements are generally that orbital energy diffusion can happen in a time less than $t_*$, that energy diffusion does not lead to the escape of most comets, and (at small $a_{\rm xp}$) that the comets do not directly impact the perturbing planet before beginning angular momentum diffusion.}
\label{fig:OortConstraints}
\end{figure}

Taken together, Eqs. \ref{constraint1}, \ref{constraint4}, and especially \ref{constraint5} provide lower limits on the needed exoplanetary mass to form an OCA, while Eqs. \ref{constraint2}, \ref{constraint3}, and \ref{constraint6} provide corresponding upper limits.  These constraints are plotted for a fiducial scenario in Fig. \ref{fig:OortConstraints}; in general it is Eq. \ref{constraint1}, Eq. \ref{constraint2}, Eq. \ref{constraint6}, and at small $a_{\rm xp}$, Eq. \ref{constraint5} that most tightly regulate the mass of the required exoplanet.  

In this section we have neglected certain additional physical effects - for example, radial migration of a star's orbit through the galactic potential \citep{Kai+11}, or dense stellar birth environments \citep{Brasse+06, KaiQui08} - in order to provide a simple analytic picture.  In general, numerical orbital integrations will be necessary to fully model the birth and evolution of OCA candidates.

\section{Impulsive Mass Loss}

When mass loss is impulsive, calculation of post-kick orbital parameters is straightforward.  Specifically, if we start with a set of orbital elements $\{a, e, i, \Omega, \omega, f\}$\footnote{The orbital elements $\{a, e, i, \Omega, \omega, f\}$ represent semimajor axis, eccentricity, inclination, longitude of ascending node, argument of pericenter, and true anomaly, respectively.}, we can obtain position and velocity coordinates using the well-known transformation rules
\begin{eqnarray}
X &=& r\left(\cos\Omega \cos (\omega+f) - \sin\Omega \sin (\omega + f) \cos i \right) \\
Y &=& r\left(\sin\Omega \cos (\omega+f) - \cos\Omega \sin (\omega + f) \cos i \right) \\
Z &=& r \sin (\omega + f) \sin i \\
\dot{X} & =& \frac{-na}{\sqrt{1-e^2}} [\cos\Omega (e\sin\omega + \sin(\omega + f) ) \\
&&+ \cos i \sin\Omega (e\cos\omega + \cos (\omega + f))] \notag \\
\dot{Y} &=& \frac{-na}{\sqrt{1-e^2}} [\sin\Omega (e\sin\omega + \sin(\omega + f) ) \\
&&- \cos i \cos\Omega (e\cos\omega + \cos (\omega + f))]\notag \\ 
\dot{Z} &=& \frac{na}{\sqrt{1-e^2}} \left( e\cos\omega\sin i + \cos (\omega+f) \sin i \right).
\end{eqnarray}
Here  the mean motion $n=\sqrt{GM_{\rm WD}/a^3}$ and the orbital radius $r=a(1-e^2)/(1+e\cos f)$.  With our orbital elements in Cartesian form, we now incorporate an impulsive kick by defining $\dot{Z}' = \dot{Z} + v_{\rm k}$, accounting for stellar mass loss by defining $M'_{\rm WD}=f_{\rm loss}M_{\rm WD}$, and recalculating standard orbital elements using the post-kick specific angular momentum vector $\vec{h}'$:
\begin{eqnarray}
a'&=&\left( \frac{2}{r} - \frac{v'^2}{GM'_{\rm WD}} \right) \\
e'&=&\sqrt{1-\frac{h'^2}{GM'_{\rm WD}a'}} \\
i'&=&\cos ^{-1} \left(\frac{h'_{\rm Z}}{h'} \right) \\
\sin\Omega ' &=& \frac{\pm h'_{\rm X}}{h' \sin i'},\\
cos\Omega ' &=& \frac{\mp h'_{\rm Y}}{h' \sin i'} \\
\sin (\omega ' + f') &=& \frac{Z}{R\sin i'}, \\
\cos (\omega ' + f') &=& \sec\Omega ' \left( \frac{X}{r} + \sin\Omega' \sin (\omega' + f') \cos i' \right) \nonumber \\
\\
\sin f' &=& \frac{a'(1-e'^2)}{h'e'}\dot{r}', \\
\cos f' &=& \frac{1}{e'} \left( \frac{a'(1-e'^2)}{r} -1 \right).
\end{eqnarray}

\section{Evaporative Mass Loss Over Multiple Orbits}

The case of sublimation over multiple pericenter passages can be treated as a succession of partial sublimations which each remove a fraction $t_{\rm p}/t_{\rm ev}$ of the cometary mass.  On each subsequent passage, however, the WD luminosity has decreased somewhat $L_{\rm WD} \propto t^{-\lambda}$ where $\lambda \simeq 5/4$ (Eq.~[\ref{eq:LWD}]), such that the mass loss per passage is reduced.  The total fractional mass lost by a single comet over $N$ orbits is therefore given by
\begin{equation}
\frac{\Delta M_{N}}{M_{\rm c}} = \sum_i^N \frac{t_{\rm p}}{t_{\rm ev}} = \sum_i^N 2.2 R_{\rm c, km}^{-1} R_{\rm p, AU}^{-1/2} M_{0.6}^{(\lambda-1)/2} a_{2000}^{-3\lambda/2} i^{-\lambda},
\label{eq:sum}
\end{equation}
where the comet's semimajor axis $a$ has been normalized as $a_{2000}=a/(2000~{\rm AU})$.  The true mass fraction lost is the smaller of $\Delta M_{N}/M_{\rm c}$ and 1.  

Over an infinite number of orbits, the sum (\ref{eq:sum}) will converge provided $\lambda > 1$, in which case the sublimated mass is given by
\begin{equation}
\frac{\Delta M_{\infty}}{M_{\rm c}} =2.2 R_{\rm c, km}^{-1} R_{\rm p, AU}^{-1/2} M_{0.6}^{(\lambda-1)/2} a_{2000}^{-3\lambda/2} \zeta ( \lambda), 
\end{equation}
where $\zeta$ is the Riemann zeta function.  For our fiducial case, $\zeta(1.25)=4.6$, implying that multiple pericenter passages can increase the maximum mass sublimated from larger comets by a factor of a few.  Only if $\lambda$ is fine-tuned to be close to 1 will multiple pericenter passages have an order-of-magnitude or greater impact on mass loss.  Also note that only comets on bound orbits will have multiple pericenter passages; sublimation over multiple orbits is therefore largely irrelevant for large kicks $v_{\rm k} \gtrsim 2~{\rm km~s}^{-1}$ that place most comets on hyperbolic orbits.  

\section{Eccentric Poynting-Robertson Drag}

The solid-state debris left over after sublimation of the volatiles in a comet will travel on highly eccentric orbits, which are subject to Poynting-Robertson drag from the central WD of luminosity $L_{\rm WD}$.  Here we outline the dynamics of PR drag on dust grains of size $b$ for general values of the orbital eccentricity.

The secular changes in orbital semimajor axis $a$ and eccentricity $e$ due to PR drag are given in \citet{Burns+79} as
\begin{align}
\left \langle \frac{{\rm d}a}{{\rm d}t} \right \rangle = -\frac{\eta Q_{\rm PR}}{a} \frac{2+3e^2}{(1-e^2)^{3/2}} \label{eq:da}\\
\left \langle \frac{{\rm d}e}{{\rm d}t} \right \rangle = -\frac{5\eta Q_{\rm PR}}{2a^2} \frac{e}{(1-e^2)^{1/2}} \label{eq:de},
\end{align}
where 
\begin{equation}
\eta = \frac{3L_{\rm WD}}{4b \rho_{\rm d} c^2},
\end{equation}
and $Q_{\rm PR}$ is a dimensionless transmission efficiency coefficient that incorporates the effects of absorption and scattering; we will focus on the limit in which the wavelength of the incident radiation satisfies $\lambda \ll b$.

\begin{figure}
\includegraphics[width=85mm]{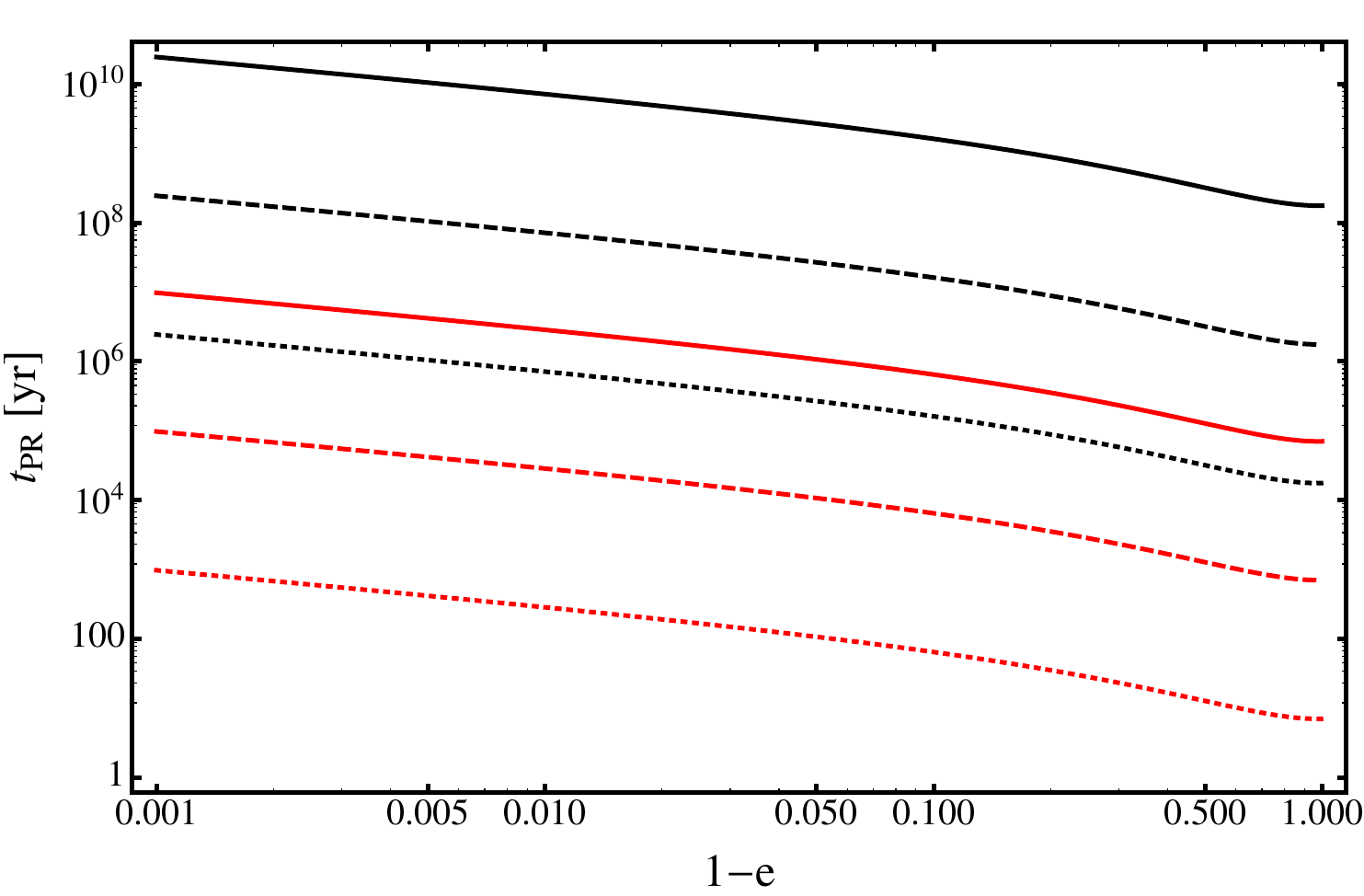}
\caption{Poynting-Robertson drag timescale, $t_{\rm PR}$, as a function of the eccentricity deficit $1-e$ of the initial orbit, calculated for a WD of luminosity $L_{\rm WD} = 100L_{\odot}$ and for orbits with different pericenter radii $r_{\rm p}=50~{\rm AU}$ ({\it black}) and 1 AU ({\it red}).  Solid, dashed, and dotted lines are calculated for particles of size $b = 10^4~{\rm \mu m}$, $10^2~{\rm \mu m}$, and $1~{\rm \mu m}$, respectively.  Note that $t_{\rm PR}$ represents only the `initial' drag timescale, which will in general increase ($\propto L_{\rm WD}^{-1} \propto t^{1.2}$, approximately; Eq.~\ref{eq:LWD}) as the WD cools, such that the actual drag time becomes infinite once $t_{\rm PR}$ exceeds the WD cooling timescale.  The effect of eccentricity for values $1-e \sim 10^{-1}-10^{-2}$ (typical of the OCA debris streams of interest here) is to increase the drag timescale by up to an order of magnitude as compared to the circular case (right edge of the plot).}
\label{fig:TPR}
\end{figure}

By combining equations \eqref{eq:da} and \eqref{eq:de}, the semimajor axis and eccentricity can be integrated from their initial values $\{a_0, e_0 \}$ to final values $\{a, e\}$:
\begin{equation}
\frac{a}{a_0} = \left(\frac{e}{e_0} \right)^{4/5} \frac{1-e_0^2}{1-e^2}. \label{eq:aOfe}
\end{equation}
Thus we observed that for the semimajor axis or pericenter radius of a particle to reach zero, the eccentricity must also go to zero.  By substituting equation \eqref{eq:aOfe} back into (\ref{eq:de}) and integrating, one obtains an implict expression for the PR drag timescale $t_{\rm PR}$ generalized to eccentric orbits:
\begin{equation}
\int_{e_0}^0 e^{3/5}(1-e^2)^{-3/2}{\rm d}e = \int_0^{t_{\rm PR}} -\frac{5 \eta Q_{\rm PR}}{2a_0^2} \frac{e_0^{8/5}}{(1-e_0^2)^{2}} {\rm d}t. \label{eq:eIntegral}
\end{equation}
Though closed form expressions for $t_{\rm PR}$ exist for general $e_0$ (using hypergeometric functions), for our purposes we specialize to the $e_0 \approx 1$ limit, in which case
\begin{align}
t_{\rm PR} \underset{e \approx 1}\approx & \frac{a_0^2(1-e_0^2)^2}{20 \eta Q_{\rm PR}} \left( \frac{8}{(1-e_0^2)^{1/2}} - \frac{3\pi^{1/2} \Gamma (9/5)}{\Gamma( 13/10)}\right) \\ 
 \approx & \frac{4\sqrt{2}a_0^{1/2}r_{\rm p, 0}^{3/2}}{5 \eta Q_{\rm PR}}, \notag 
\end{align}
where $\Gamma$ is the Gamma function and $r_{\rm p, 0}$ is the initial orbital pericenter.  In the main text, we assume $Q_{\rm PR}=1$.  Figure \ref{fig:TPR} shows the PR timescale $t_{\rm PR}$ as a function of initial eccentricity, as obtained from an exact solution to equation \eqref{eq:eIntegral}.

\end{document}